\documentclass[aps,prl,twocolumn,groupedaddress,superscriptaddress,nobibnotes]{revtex4-2}
\usepackage{graphicx}
\usepackage{epstopdf}
\usepackage{amssymb, amsmath}
\usepackage[usenames]{color}
\usepackage{bm}
\usepackage{ulem}
\usepackage{multirow}
\usepackage{threeparttable}
\usepackage{booktabs}
\usepackage{dcolumn}
\usepackage{float}
\usepackage{hyperref}

\hypersetup{
	colorlinks=true,
	linkcolor=blue,
	filecolor=gray,      
	urlcolor=blue,
	citecolor=blue,
}

\begin{document}

\title{High-Chern-number Quantum anomalous Hall insulators in mixing-stacked MnBi$_2$Te$_4$ thin films}

\author{Jiaheng Li}
\affiliation{Beijing National Laboratory for Condensed Matter Physics and Institute of Physics, Chinese Academy of Sciences, Beijing 100190, China}

\author{Quansheng Wu}
\email{quansheng.wu@iphy.ac.cn}
\affiliation{Beijing National Laboratory for Condensed Matter Physics and Institute of Physics, Chinese Academy of Sciences, Beijing 100190, China}
\affiliation{University of Chinese Academy of Sciences, Beijing 100049, China}

\author{Hongming Weng}
\email{hmweng@iphy.ac.cn}
\affiliation{Beijing National Laboratory for Condensed Matter Physics and Institute of Physics, Chinese Academy of Sciences, Beijing 100190, China}
\affiliation{University of Chinese Academy of Sciences, Beijing 100049, China}
\affiliation{Songshan Lake Materials Laboratory, Dongguan, Guangdong 523808, China}

\begin{abstract}
Quantum anomalous Hall (QAH) insulators are characterized by vanishing longitudinal resistance and quantized Hall resistance in the absence of an external magnetic field. Among them, high-Chern-number QAH insulators offer multiple nondissipative current channels, making them crucial for the development of low-power-consumption electronics. Using first-principles calculations, we propose that high-Chern-number ($C>1$) QAH insulators can be realized in MnBi$_2$Te$_4$ (MBT) multilayer films through the combination of mixed stacking orders, eliminating the need for additional buffer layers. The underlying physical mechanism is validated by calculating real-space-resolved anomalous Hall conductivity (AHC). Local AHC is found to be predominantly located in regions with consecutive correct stacking orders, contributing to quasi-quantized AHC. Conversely, regions with consecutive incorrect stacking contribute minimally to the total AHC, which can be attributed to the varied interlayer coupling in different stacking configurations. Our work provides valuable insights into the design principle for achieving large Chern numbers, and highlights the role of stacking configurations in manipulating electronic and topological properties in MBT films and its derivatives.
\end{abstract}

\maketitle

Quantum anomalous Hall (QAH) states, as the typical two-dimensional topological insulating state, exhibit vanishing longitudinal resistance, and quantized Hall resistance in the absence of an external magnetic field\cite{haldane1988model}, in which dissipationless chiral edge states encircling QAH insulators (QAHIs) are anticipated to be advantageous in advancing low-power electronics. There have been several experimental observations of QAHIs with $C=1$ at low temperatures, such as magnetically doped topological insulator (TI) films and twist moiré systems \cite{chang2013experimental, checkelsky2014trajectory, chang2015zero, mogi2017magnetic, ou2018enhancing, deng2020quantum, serlin2020intrinsic, li2021quantum}. Although chiral edge current exists QAHIs with $C=1$, the presence of contact resistance still constrains these proof-of-concept devices to have limited breakdown currents. High-Chern-number QAHIs provide more nondissipative chiral current channels and decrease the effective Hall resistance of $h/Ce^2$, facilitating the practical applications of the QAH devices. Despite there have been a series of theoretical predictions on high-Chern-number QAHIs in magnetically doped TIs under strong exchange field \cite{wang2013quantum} and electric field \cite{du2020berry}, heterostructures spaced by additional buffer layers \cite{burkov2011weyl, bosnar2023high}, quantized Hall conductance has predominantly been observed in magnetically doped topological insulator films \cite{zhao2020tuning}. Thus, there remains a pressing demand for discovering and engineering more high-Chern-number QAHIs in realistic material systems \cite{chang2023quantum}.

Meanwhile, extensive theoretical predictions \cite{li2019intrinsic, zhang2019topological, otrokov2019unique} and experimental fabrication \cite{lee2013crystal, gong2019experimental, yan2019crystal, cui2019transport} have identified MnBi$_2$Te$_4$ (MBT) as an excellent candidate for exploring the fascinating interplay among spatial dimension, topology and magnetism\cite{li2019intrinsic, otrokov2019unique, li2019magnetically, liu2020robust, deng2020quantum}. This material has demonstrated a wide range of extraordinary physical phenomena, including mysterious Dirac cones at the top surface \cite{chen2019topological, otrokov2019prediction}, QAH insulators \cite{deng2020quantum}, axion insulating states \cite{liu2020robust, zhu2021tunable},and high-Chern-number Chern insulators under external magnetic field \cite{ge2020high}. Notably, as a typical van der Waals material, interlayer coupling strength in MBT multi-layer thin films undergoes significant variations under distinct stacking orders, even giving rise to stacking-order dependent topology \cite{ren2022quantum, zhu2022high, cao2023switchable, ahn2023stacking, li2025stacking, wang2023three}. Given the unique properties of MBT and the pressing demand for high-Chern-number QAH insulators, further exploration of MBT and its derivatives holds great promise for overcoming current limitations and advancing the field of quantum topological materials.

\begin{figure}[htbp]
	\includegraphics[width=1.0\linewidth]{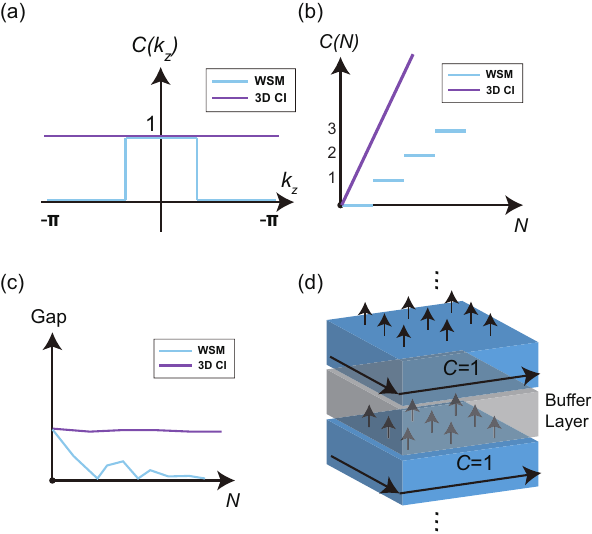}
	\caption{Schematic illustration of electronic and topological properties in the bulk and slab from typical magnetic Weyl semimetals (WSMs) and 3D Chern insulators(CIs). (\textbf{a}) Chern number defined on a series of $k_z$ planes. (\textbf{b}) Chern numbers and (\textbf{c}) band gaps of slabs cut from bulk WSM and 3D CI as a function of layer number ($N$). (\textbf{d}) Typical 3D CIs built from periodic pattern of 2D Chern insulators with buffer layers along the stacking direction.}
	\label{fig1}
\end{figure}

The realization of QAHIs with $C>1$ can be achieved through two distinct strategies: the design of novel functional two-dimensional (2D) materials and the construction of heterostructures by stacking van der Waals (vdW) layers from bulk counterparts. Among these approaches, the most convenient way is to cut slabs from three-dimensional (3D) Chern insulators or Weyl semimetals (WSMs) with the vdW nature. Notably, Chern numbers can be defined on all closed 2D manifolds, such as 2D Brillouin zone and the sphere surrounding Weyl points. As shown in Fig. \ref{fig1} (a), the simplest 3D Chern insulator can be regarded as the periodic pattern of 2D Chern insulators in the stacking direction with negligible interlayer coupling due to the existence of buffer layer. This 3D Chern insulator possesses consistent $C=1$ on these $k_z$ plane (Fig. \ref{fig1} (b)), but Chern number in WSMs abruptly jumps when $k_z$ planes cross the Weyl point. Chern numbers of slabs cut from 3D Chern insulators can exhibit a linear increase as the number of layers is incremented, and band gaps of thin films almost keep constant (Fig. \ref{fig1}(c)). Conversely, the Chern number of thin films derived from WSMs shows a step-like increase, and exhibits an oscillating gap with an overall decreasing trend as the number of layers increases. Nevertheless, the materialization of 3D vdW Chern insulators, which inherently possess the vdW characteristics, remains elusive. This significant gap impedes both the material realization and experimental observation of QAH insulators with high Chern numbers.

In the work, we demonstrate that MBT thin films with mixed stacking order can achieve arbitrarily large Chern numbers without introducing additional buffer layer (Fig. \ref{fig1}(d)), which originates from varied interlayer coupling present in the different stacking configurations. Within these QAHIs, local anomalous Hall conductivity (AHC) is found to be concentrated in the consecutive correct stacking order, while the contribution from the consecutive incorrect stacking order is negligible. Furthermore, the bulk gaps in these QAHIs can be tuned to be over 20 meV by increasing layer numbers of both correct and incorrect stacking orders, through utilizing first-principles calculations and tight-binding effective Hamiltonians.

\begin{figure}[htbp]
	\centering
	\includegraphics[width=0.9\linewidth]{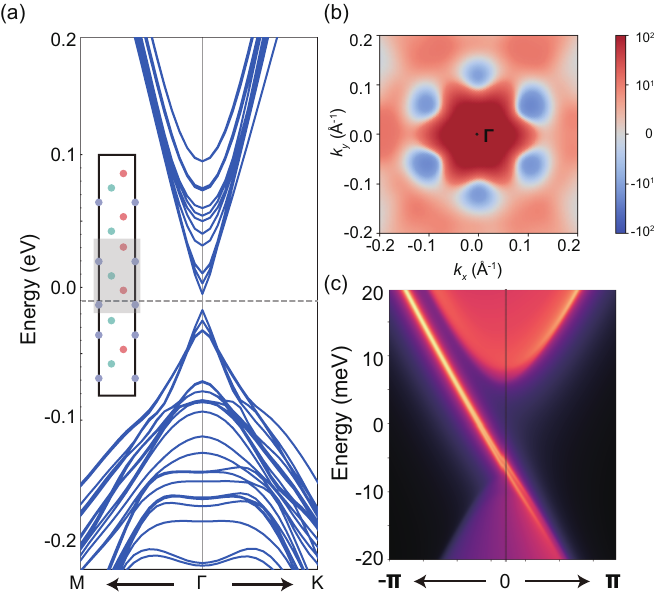}
	\caption{Electronic and topological properties of 15-SL MBT thin films with stacking order in the ABCABACBACBCABC sequence. (\textbf{a}) Electronic band structures along high symmetry line $M-\Gamma-K$. (\textbf{b}) Distribution of Berry curvatures in the Brillouin zone. (\textbf{c}) Local density of states on the [110] edge.}
	\label{fig2}
\end{figure}

\section{Results}
\subsection{First-principles calculations}
Bulk MBT crystalline in the $R\bar{3}m$ structure, which is characterized by the ABC-stacked building blocks in the Te-Bi-Te-Mn-Te-Bi-Te sequence \cite{lee2013crystal, li2019intrinsic, gong2019experimental, yan2019crystal, cui2019transport}. Interlayer sliding can be realized in MBT thin films due to weak vdW interactions between neighboring SLs, allowing for different stacking orders, such as AB and AC high-symmetry stacking order. The most stable stacking order is AB-stacking, which exactly corresponds to the normal bulk form, while AC-stacking is energetically higher by 4 meV per atom, compared to AB-stacking. Dynamical stability is validated by phonon dispersion without imaginary phonon modes \cite{SM}. Therefore, thick films with mixed stacking orders are expected to be dynamically stable, and can be realized in the experiments. 

In the AB-stacked bulk, the interlayer coupling is more pronounced due to the reduced interlayer spacing of 2.8 $\mathrm{\AA}$, in contrast to 3.2 $\mathrm{\AA}$ observed in the AC-stacking. Stronger interlayer coupling, combined with significant spin-orbit coupling from the Bi and Te atoms, leads to the emergence of a topological band inversion at $\Gamma$ in the AB-stacked configuration. Consequently, AB-stacked AFM bulk belongs to antiferromagnetic (AFM) TIs with $Z_2=1$ \cite{li2019intrinsic, otrokov2019unique, zhang2019topological}, and the AC-stacked AFM bulk falls in a topological trivial insulating case with $Z_2=0$, elucidating the strong coupling in stacking order and topology in MBT materials \cite{li2025stacking}. When MBT bulks is exfoliated into films, the consecutive ABC-stacking order exhibit odd-even oscillation between QAH insulators and axion insulators \cite{li2019intrinsic, otrokov2019unique}, while their counterparts with the consecutive CBA-stacking order persist as unequivocal trivial magnetic insulators without any topological feature. Below, both stacking orders are simultaneously employed to fabricate QAHIs endowed with large Chern numbers ($C>1$), where the consecutive ABC stacking order is utilized to form QAH insulating layers with $C=1$, and the CBA stacking order is employed as the buffer layer with $C=0$.

As a representative example, let us consider 15-SL films with the ABCABACBACBCABC stacking sequence, shown in the inset of Fig. \ref{fig2} (a). According to varied interlayer coupling strength, the entire thin film can be divided into consecutive correct stacking layers with the ...ABC... stacking order and consecutive incorrect stacking layers with the ...CBA... stacking order. Thus, the above 15-SL film can be divided into the ABCAB-ACBAC-BCABC stacking pattern, represented as $(5, -5, 5)$. This representation indicates that this 15-SL film begins with five consecutive correct stacking layers (ABC), followed by five consecutive incorrect stacking layers (CBA), and ends with another set of five consecutive correct stacking layers (ABC). This alternating sequence of consecutive correct and incorrect stacking layers is essential for inducing large Chern number QAH insulators.  

Electronic band structures of the $(5, -5, 5)$ stacking configuration along these high-symmetry lines $M-\Gamma-K$ are shown in Fig. \ref{fig2}(a), where a direct band gap of 18 meV is located at $\Gamma$. Furthermore, the gapped Dirac cone observed in this band structure resembles the gapped Dirac cones on the top surface of ABC-stacked MBT bulk, implying that films inherit crucial electronic and topological properties of AFM TIs, despite their reduced dimensionality. The $k$-resolved Berry curvature in Fig. \ref{fig2}(b) shows that Berry curvatures are mainly concentrated around $\Gamma$, and exhibit sixfold rotational symmetry, guaranteed by the presence of spatial inversion symmetry $\mathcal{I}$ and three-fold rotational symmetry $C_{3z}$. The sum of Berry curvature in the whole Brillouin zone confirms that (5, -5, 5) film possesses $C=2$. Furthermore, the calculated local density of states on the semi-infinite ribbon in Fig. \ref{fig2}(c) clearly reveals the presence of two chiral edge modes that transverse the bulk band gap.

To further illustrate the layer-resolved AHC in 15-SL MBT films, a numerical approach based on local basis sets has been implemented to calculate the real-space Chern-Simons contribution to AHC \cite{bianco2011mapping, marrazzo2017locality, varnava2018surface},

\begin{equation}
	C(\alpha) = -{4 \pi}\mathrm{Im}\int_{\mathrm{BZ}}d\mathbf{k} \sum_{vv'c}\psi_{v\mathbf{k}}^\alpha \mathbf{X}_{vc\mathbf{k}} \mathbf{Y}^\dagger_{v'c\mathbf{k}}\psi_{v'\mathbf{k}}^{\alpha*}, 
\end{equation}
where 
\begin {equation}
\mathbf{X}_{vc\mathbf{k}} = \langle \psi_{v\mathbf{k}} | \hat{x} | \psi_{c\mathbf{k}} \rangle = \frac{\langle \psi_{v\mathbf{k}} | i\hbar \hat{v}_x | \psi_{c\mathbf{k}} \rangle }{E_{c\mathbf{k}} - E_{v\mathbf{k}}}, 
\end{equation}
and $Y_{vc\mathbf{k}}$ take the similar form. Here $|\psi_{v\mathbf{k}}\rangle$ ($E_{v\mathbf{k}}$) and $|\psi_{c\mathbf{k}}\rangle$ ($E_{c\mathbf{k}}$) represent eigenstates (eigenvalues) in the valence and conduction bands with $\mathbf{k}= (k_x, k_y)$ traversing the entire Brillouin zone, respectively. $\hat{v}_x$ and $\hat{v}_y$ are the velocity operator defined as $\frac{1}{\hbar}\frac{\partial \hat{H}}{\partial \mathbf{k}}$, and the symbol $\alpha$ denotes local basis set. And layer-dependent Chern number $C(l)$ can be defined as the sum of real-space AHC terms of basis sets located with this layer ($l$),
\begin{equation}
    C(l) = \sum_\alpha C(\alpha).
\end{equation}

\begin{figure}[htbp]
	\centering
	\includegraphics[width=1.0\linewidth]{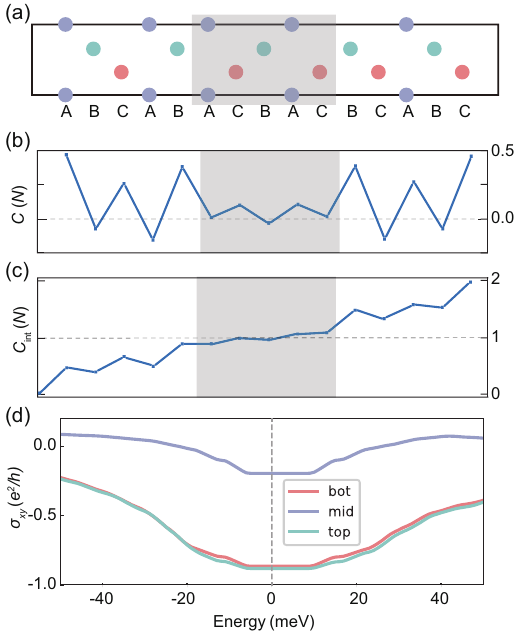}
	\caption{Local anomalous Hall conductivity (AHC) in the 15-SL MBT thin films with ABCABACBACBCABC (5, -5, 5) stacking order. (\textbf{a}) Schematic illustration of its crystal structure, with each consecutive stacking order marked by colored shades. (\textbf{b}) Calculated local AHC of each layer $C(N)$ and (\textbf{c}) the cumulative sum of local AHC $C_{\mathrm{int}}(N)= \sum_{i=1}^N C(i)$. (\textbf{d}) The sum of local AHC in the consecutive correct (incorrect) stacking order as a function of chemical potential shift ($E_F$), where the top, middle and bottom 5-SL are denoted by \texttt{bot}, \texttt{mid} and \texttt{top}, respectively.}
	\label{fig3}
\end{figure}

\subsection{Comparison between effective Hamiltonians and tight-binding models}
Based on the above equation, detailed calculations have been performed to determine the layer-resolved AHC in (5, -5, 5) thin films, as shown in Fig. \ref{fig3}(a). The oscillation of local AHC between neighboring SLs in Fig. \ref{fig3}(b) is attributed to an effective Zeeman field induced by opposite magnetic moments located on Mn atoms, originating from the A-type antiferromagnetic configuration. The top and bottom SLs of the consecutive correct stacking layers exhibit the largest and nearly half-quantized AHC (Fig. \ref{fig3}(b)), which is indeed the signature of half-quantized AHC on the top surface in the ABC-stacking bulk phase. The cumulative sum of layer-resolved AHC is presented in Fig. \ref{fig3}(c), and its increment reveals noticeable discrepancy across different stacking regions. Notably, the local AHC is predominantly concentrated in the top and bottom regions consisting of consecutive correct stacking order. In contrast, the middle region, which consists of incorrect stacking layers, contributes little to the overall AHC. When summing the contributions of AHC from these three consecutive stacking regions, we observe that both top and bottom consecutive regions exhibit a quasi-quantized AHC behavior (Fig. \ref{fig3} (d)). The layer-resolved contribution to total AHC not only provides valuable insights into the underlying physics of topological electronic states, but also is essential for unraveling the mechanisms that govern AHC and for optimizing the performance of QAH insulators.


To provide a quantitatively good description of electronic and topological properties in multi-SL MBT films of mixed AB- and AC-stacking orders, a low-energy effective model Hamiltonian is constructed by adding the onsite Zeeman term into the widely adopted Bernevig-Hughes-Zhang (BHZ) model \cite{bernevig2006quantum, zhang2009topological, liu2010model}. Therefore, the effective model Hamiltonians, that describe varied interlayer coupling in distinct stacking configurations, can be written as

\begin{equation}
	\begin{aligned}
		H(\mathbf{k}_\parallel) = &\sum_{\mathbf{k}, ij}\left[(H_{0}(\mathbf{k})+ H_Z)\delta_{i, j} \right.\\
		&+ \left. H_+(\mathbf{k})\delta_{i,j-1} + H_-(\mathbf{k}) \delta_{i, j+1}\right],
	\end{aligned}
\end{equation}
where $i$ denotes the $i$th SL in the films, $H_0$, $H_{+}$ ($H_{-}$), and $H_Z$ represent the intralayer Hamiltonian, interlayer coupling and onsite Zeeman term, respectively. $\tau$ and $\sigma$ operate on the orbital and spin degrees of freedom. The latter two terms are assumed to be dependent only on stacking order. Following the effective $\vec{\mathbf{k}}\cdot\vec{\mathbf{p}}$ Hamiltonians based on four low-lying state $|\mathrm{Bi}_z^{+}, \uparrow(\downarrow)\rangle$ and $|\mathrm{Te}_z^{-}, \uparrow(\downarrow)\rangle$, the intralayer and interlayer Hamiltonians can be written as 
\begin{equation}
	\begin{aligned}
		H_{0} &= C \tau_0 \sigma_0 + M \tau_z\sigma_0 + A(k_x \tau_x \sigma_x + k_y \tau_x \sigma_y), \\
		H_{+} &= -C_1 \tau_0 \sigma_0 - M_1 \tau_z \sigma_0 + i B_0 \tau_x \sigma_z, \\
		H_{Z} &= Z_1 \tau_0 \sigma_z + Z_2 \tau_z \sigma_z, 
	\end{aligned}
\end{equation}

where $C = C_0 + C_2 (k_x^2 + k_y^2)$, $M= M_0 + M_2 (k_x^2 + k_y^2)$, and all the higher-order hopping parameters are ignored. $H_{-} = H_{+}^\dagger$, which correspond to interlayer hopping terms to the lower and upper neighboring SL, respectively. By incorporating an onsite Zeeman term into the BHZ model, the essential low-energy physics can be effectively captured, allowing for a quantitative depiction of both electronic and topological properties.

\begin{figure}[htbp]
	\centering
	\includegraphics[width=1.0\linewidth]{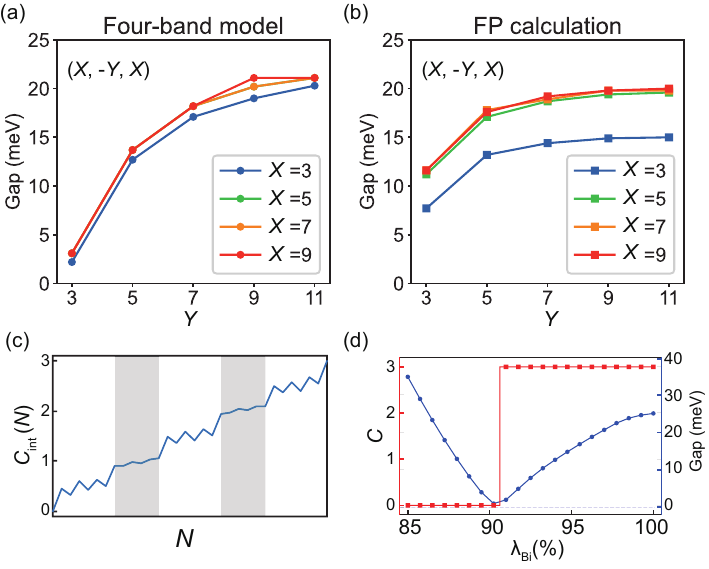}
	\caption{Band gaps of distinct stacking patterns $(X, -Y, X)$ calculated by four-band effective model (\textbf{a}) and first-principles calculations (\textbf{b}). (\textbf{c}) The cumulative sum of Chern number $C(N)$ as a function of layer number, where the consecutive correct (incorrect) stacking part are shaded by white (gray). (\textbf{d}) Chern number ($C$) and band gap as a function of artificial spin-orbit coupling strength $\lambda_{\mathrm{Bi}}$ in the $(7, -5, 7, -5, 7)$ stacking configuration.}
	\label{fig4}
\end{figure}

By conducting first-principles calculations in distinct mixed stacking patterns and employing the aforementioned effective model Hamiltonian, we illustrate the evolution of band gaps across various stacking order scenarios characterized by the sequence $(X, -Y, X)$ in Fig. \ref{fig4} (a) and (b). Our effective model Hamiltonians are in good agreement with first-principles calculations, demonstrating the reliability and accuracy of the effective model in the multi-SL MBT films with mixed stacking configurations. All these stacking configurations yield large Chern number QAHIs with $C=2$. Similar to the former calculation of local AHC in the (5, -5, 5) film, the consecutive incorrect stacking layers with odd SLs ($Y \ge 3$) can be regarded as buffer layers that separate periodic patterned QAHIs in the stacking direction and strongly suppress the hybridization of QAH insulating layers. Contrarily, the consecutive correct stacking layers with odd SLs ($X \ge 3$) can be treated as an effective Chern layer with local AHC of approximately $e^2/h$. To construct high-Chern-number QAH insulators using MBT SLs, consecutive correct and incorrect regions, each with a minimum thickness of 3 SLs, should be alternated to form complete stacking configurations.

As the layer number of consecutive correct and incorrect stacking configurations increases, band gaps demonstrate a clear increasing trend from several meV to tens of meV, highlighting a clear tendency towards larger band gaps in thicker films, with the maximum observed band gap reaching approximately 20 meV. Thus, large Chern number with $C=N$ can be realized in $([X, -Y]..._{N-1}, X)$ multi-SL MBT films without the need for additional materials, greatly enhancing the feasibility of experimentally observing large-gap QAHIs with $C>1$.

\subsection{Doping effects}

Considering the typical electron-doped nature of MBT \cite{chen2019topological, otrokov2019prediction, li2019dirac, hao2019gapless}, experimental synthesis of Mn(Bi$_{x}$Sb$_{1-x}$)$_2$Te$_4$ (MBST) has been successfully accomplished \cite{chen2019intrinsic}, revealing a discernibly lower concentration of electron doping. Moreover, the electronic and topological characteristics of MBST can be adroitly manipulated through the variable composition parameter $x$. In order to make theoretical predictions regarding doped MBST systems, first-principles calculations that incorporate diverse atomic spin-orbit coupling ($\lambda_{\mathrm{Bi}}$) specifically on Bi element are performed to circumvent the complexity of doped systems, where $\lambda_{\mathrm{Bi}}=1.0 (0.0)$ refers to signifies the presence (absence) of normal spin-orbit coupling (SOC) on Bi element in the system. All the possible interlayer and intralayer local stacking orders are iterated, and then corresponding tight-binding Hamiltonians based on localized Wannier functions are taken as building blocks to construct arbitrary stacking configurations with varied spin-orbit coupling in thick films. 

Utilizing the aforementioned methodology, we construct 31-SL MBST film with (7, -5, 7, -5, 7) stacking consequence of varied spin-orbit coupling strength, and confirm MBST with $\lambda_{\mathrm{Bi}}=1.0$ to be QAHIs with $C=3$ and band gaps of 20 meV. The cumulative sum of layer-resolved AHC in MBT films shows behavior similar to the (5, -5, 5) case. When $\lambda_{\mathrm{Bi}}$ is tuned from $100\%$ to $80\%$, the MBST film undergoes a topological phase transition from $C=3$ to $C=0$ at $\lambda_{\mathrm{Bi}}=0.9$. Thus, this method serves as a computationally efficient tool for investigating the electronic and topological properties of MBST films with arbitrary stacking configurations, without requiring extensive and time-consuming first-principles calculations.

\section{Discussion}
Our work predicts a series of high-Chern-number quantum anomalous Hall insulators (QAHIs) with $C>1$ in MnBi$_2$Te$_4$ (MBT) multilayer thin films, highlighting the significant role of mixed stacking orders in achieving large Chern numbers. We demonstrate the feasibility of constructing such QAHIs in typical van der Waals magnetic topological materials, obviating the necessity for extraneous interstitial buffer layers. Our further calculations on real-space-resolved anomalous Hall conductivity (AHC) within these QAHIs reveal that consecutive correct stacking orders are primarily responsible for the total AHC, while consecutive incorrect stacking orders contribute negligibly. Moreover, our combined approach of utilizing first-principles calculations and tight-binding effective Hamiltonians proves to be a highly powerful and accurate tool for quantitatively predicting electronic and topological properties of MBT heterostructures with mixed stacking orders. Our results offer strategic ramifications for the advancement of QAHIs with enhanced performance in next-generation topological electronic devices.

\section{Acknowledgments}
This work was supported by the National Key R\&D Program of China (Grant No. 2022YFA1403800, 2023YFA1607401), the National Natural Science Foundation of China (Grant No. 12274436, 11925408, 11921004), the Science Center of the National Natural Science Foundation of China (Grant No. 12188101) and  H.W. acknowledge support from the Informatization Plan of the Chinese Academy of Sciences (CASWX2021SF-0102) and the New Cornerstone Science Foundation through the XPLORER PRIZE. J. L. acknowledge support from China National Postdoctoral Program for Innovation Talents (Grant No. BX20220334).

\section{Contributions}
J.L., Q.W., and H.W. conceived the project. J. L. performed first-principles calculations, and data analysis and wrote the initial manuscript with the help of Q.W., and H.W. All authors were engaged in discussing the main results and editing the final manuscript.

\hypersetup{hidelinks=true}
\bibliographystyle{apsrev4-2}

\begin{thebibliography}{44}%
	\makeatletter
	\providecommand \@ifxundefined [1]{%
		\@ifx{#1\undefined}
	}%
	\providecommand \@ifnum [1]{%
		\ifnum #1\expandafter \@firstoftwo
		\else \expandafter \@secondoftwo
		\fi
	}%
	\providecommand \@ifx [1]{%
		\ifx #1\expandafter \@firstoftwo
		\else \expandafter \@secondoftwo
		\fi
	}%
	\providecommand \natexlab [1]{#1}%
	\providecommand \enquote  [1]{``#1''}%
	\providecommand \bibnamefont  [1]{#1}%
	\providecommand \bibfnamefont [1]{#1}%
	\providecommand \citenamefont [1]{#1}%
	\providecommand \href@noop [0]{\@secondoftwo}%
	\providecommand \href [0]{\begingroup \@sanitize@url \@href}%
	\providecommand \@href[1]{\@@startlink{#1}\@@href}%
	\providecommand \@@href[1]{\endgroup#1\@@endlink}%
	\providecommand \@sanitize@url [0]{\catcode `\\12\catcode `\$12\catcode `\&12\catcode `\#12\catcode `\^12\catcode `\_12\catcode `\%12\relax}%
	\providecommand \@@startlink[1]{}%
	\providecommand \@@endlink[0]{}%
	\providecommand \url  [0]{\begingroup\@sanitize@url \@url }%
	\providecommand \@url [1]{\endgroup\@href {#1}{\urlprefix }}%
	\providecommand \urlprefix  [0]{URL }%
	\providecommand \Eprint [0]{\href }%
	\providecommand \doibase [0]{https://doi.org/}%
	\providecommand \selectlanguage [0]{\@gobble}%
	\providecommand \bibinfo  [0]{\@secondoftwo}%
	\providecommand \bibfield  [0]{\@secondoftwo}%
	\providecommand \translation [1]{[#1]}%
	\providecommand \BibitemOpen [0]{}%
	\providecommand \bibitemStop [0]{}%
	\providecommand \bibitemNoStop [0]{.\EOS\space}%
	\providecommand \EOS [0]{\spacefactor3000\relax}%
	\providecommand \BibitemShut  [1]{\csname bibitem#1\endcsname}%
	\let\auto@bib@innerbib\@empty
	\bibitem [{\citenamefont {Haldane}(1988)}]{haldane1988model}%
	\BibitemOpen
	\bibfield  {author} {\bibinfo {author} {\bibfnamefont {F.~D.~M.}\ \bibnamefont {Haldane}},\ }\bibfield  {title} {\bibinfo {title} {Model for a quantum {Hall} effect without {Landau} levels: Condensed-matter realization of the "{Parity Anomaly}"},\ }\href {https://doi.org/10.1103/PhysRevLett.61.2015} {\bibfield  {journal} {\bibinfo  {journal} {Phys. Rev. Lett.}\ }\textbf {\bibinfo {volume} {61}},\ \bibinfo {pages} {2015} (\bibinfo {year} {1988})}\BibitemShut {NoStop}%
	\bibitem [{\citenamefont {Chang}\ \emph {et~al.}(2013)\citenamefont {Chang}, \citenamefont {Zhang}, \citenamefont {Feng}, \citenamefont {Shen}, \citenamefont {Zhang}, \citenamefont {Guo}, \citenamefont {Li}, \citenamefont {Ou}, \citenamefont {Wei}, \citenamefont {Wang}, \citenamefont {Ji}, \citenamefont {Feng}, \citenamefont {Ji}, \citenamefont {Chen}, \citenamefont {Jia}, \citenamefont {Dai}, \citenamefont {Fang}, \citenamefont {Zhang}, \citenamefont {He}, \citenamefont {Wang}, \citenamefont {Lu}, \citenamefont {Ma},\ and\ \citenamefont {Xue}}]{chang2013experimental}%
	\BibitemOpen
	\bibfield  {author} {\bibinfo {author} {\bibfnamefont {C.-Z.}\ \bibnamefont {Chang}}, \bibinfo {author} {\bibfnamefont {J.}~\bibnamefont {Zhang}}, \bibinfo {author} {\bibfnamefont {X.}~\bibnamefont {Feng}}, \bibinfo {author} {\bibfnamefont {J.}~\bibnamefont {Shen}}, \bibinfo {author} {\bibfnamefont {Z.}~\bibnamefont {Zhang}}, \bibinfo {author} {\bibfnamefont {M.}~\bibnamefont {Guo}}, \bibinfo {author} {\bibfnamefont {K.}~\bibnamefont {Li}}, \bibinfo {author} {\bibfnamefont {Y.}~\bibnamefont {Ou}}, \bibinfo {author} {\bibfnamefont {P.}~\bibnamefont {Wei}}, \bibinfo {author} {\bibfnamefont {L.-L.}\ \bibnamefont {Wang}}, \bibinfo {author} {\bibfnamefont {Z.-Q.}\ \bibnamefont {Ji}}, \bibinfo {author} {\bibfnamefont {Y.}~\bibnamefont {Feng}}, \bibinfo {author} {\bibfnamefont {S.}~\bibnamefont {Ji}}, \bibinfo {author} {\bibfnamefont {X.}~\bibnamefont {Chen}}, \bibinfo {author} {\bibfnamefont {J.}~\bibnamefont {Jia}}, \bibinfo {author} {\bibfnamefont {X.}~\bibnamefont {Dai}}, \bibinfo {author} {\bibfnamefont
			{Z.}~\bibnamefont {Fang}}, \bibinfo {author} {\bibfnamefont {S.-C.}\ \bibnamefont {Zhang}}, \bibinfo {author} {\bibfnamefont {K.}~\bibnamefont {He}}, \bibinfo {author} {\bibfnamefont {Y.}~\bibnamefont {Wang}}, \bibinfo {author} {\bibfnamefont {L.}~\bibnamefont {Lu}}, \bibinfo {author} {\bibfnamefont {X.-C.}\ \bibnamefont {Ma}},\ and\ \bibinfo {author} {\bibfnamefont {Q.-K.}\ \bibnamefont {Xue}},\ }\bibfield  {title} {\bibinfo {title} {Experimental observation of the quantum anomalous {Hall} effect in a magnetic topological insulator},\ }\href {https://doi.org/10.1126/science.1234414} {\bibfield  {journal} {\bibinfo  {journal} {Science}\ }\textbf {\bibinfo {volume} {340}},\ \bibinfo {pages} {167} (\bibinfo {year} {2013})}\BibitemShut {NoStop}%
	\bibitem [{\citenamefont {Checkelsky}\ \emph {et~al.}(2014)\citenamefont {Checkelsky}, \citenamefont {Yoshimi}, \citenamefont {Tsukazaki}, \citenamefont {Takahashi}, \citenamefont {Kozuka}, \citenamefont {Falson}, \citenamefont {Kawasaki},\ and\ \citenamefont {Tokura}}]{checkelsky2014trajectory}%
	\BibitemOpen
	\bibfield  {author} {\bibinfo {author} {\bibfnamefont {J.~G.}\ \bibnamefont {Checkelsky}}, \bibinfo {author} {\bibfnamefont {R.}~\bibnamefont {Yoshimi}}, \bibinfo {author} {\bibfnamefont {A.}~\bibnamefont {Tsukazaki}}, \bibinfo {author} {\bibfnamefont {K.~S.}\ \bibnamefont {Takahashi}}, \bibinfo {author} {\bibfnamefont {Y.}~\bibnamefont {Kozuka}}, \bibinfo {author} {\bibfnamefont {J.}~\bibnamefont {Falson}}, \bibinfo {author} {\bibfnamefont {M.}~\bibnamefont {Kawasaki}},\ and\ \bibinfo {author} {\bibfnamefont {Y.}~\bibnamefont {Tokura}},\ }\bibfield  {title} {\bibinfo {title} {Trajectory of the anomalous {Hall} effect towards the quantized state in a ferromagnetic topological insulator},\ }\href {https://doi.org/10.1038/nphys3053} {\bibfield  {journal} {\bibinfo  {journal} {Nat. Phys.}\ }\textbf {\bibinfo {volume} {10}},\ \bibinfo {pages} {731} (\bibinfo {year} {2014})}\BibitemShut {NoStop}%
	\bibitem [{\citenamefont {Chang}\ \emph {et~al.}(2015)\citenamefont {Chang}, \citenamefont {Zhao}, \citenamefont {Kim}, \citenamefont {Wei}, \citenamefont {Jain}, \citenamefont {Liu}, \citenamefont {Chan},\ and\ \citenamefont {Moodera}}]{chang2015zero}%
	\BibitemOpen
	\bibfield  {author} {\bibinfo {author} {\bibfnamefont {C.-Z.}\ \bibnamefont {Chang}}, \bibinfo {author} {\bibfnamefont {W.}~\bibnamefont {Zhao}}, \bibinfo {author} {\bibfnamefont {D.~Y.}\ \bibnamefont {Kim}}, \bibinfo {author} {\bibfnamefont {P.}~\bibnamefont {Wei}}, \bibinfo {author} {\bibfnamefont {J.~K.}\ \bibnamefont {Jain}}, \bibinfo {author} {\bibfnamefont {C.}~\bibnamefont {Liu}}, \bibinfo {author} {\bibfnamefont {M.~H.~W.}\ \bibnamefont {Chan}},\ and\ \bibinfo {author} {\bibfnamefont {J.~S.}\ \bibnamefont {Moodera}},\ }\bibfield  {title} {\bibinfo {title} {Zero-field dissipationless chiral edge transport and the nature of dissipation in the quantum anomalous {Hall} state},\ }\href {https://doi.org/10.1103/PhysRevLett.115.057206} {\bibfield  {journal} {\bibinfo  {journal} {Phys. Rev. Lett.}\ }\textbf {\bibinfo {volume} {115}},\ \bibinfo {pages} {057206} (\bibinfo {year} {2015})}\BibitemShut {NoStop}%
	\bibitem [{\citenamefont {Mogi}\ \emph {et~al.}(2017)\citenamefont {Mogi}, \citenamefont {Kawamura}, \citenamefont {Yoshimi}, \citenamefont {Tsukazaki}, \citenamefont {Kozuka}, \citenamefont {Shirakawa}, \citenamefont {Takahashi}, \citenamefont {Kawasaki},\ and\ \citenamefont {Tokura}}]{mogi2017magnetic}%
	\BibitemOpen
	\bibfield  {author} {\bibinfo {author} {\bibfnamefont {M.}~\bibnamefont {Mogi}}, \bibinfo {author} {\bibfnamefont {M.}~\bibnamefont {Kawamura}}, \bibinfo {author} {\bibfnamefont {R.}~\bibnamefont {Yoshimi}}, \bibinfo {author} {\bibfnamefont {A.}~\bibnamefont {Tsukazaki}}, \bibinfo {author} {\bibfnamefont {Y.}~\bibnamefont {Kozuka}}, \bibinfo {author} {\bibfnamefont {N.}~\bibnamefont {Shirakawa}}, \bibinfo {author} {\bibfnamefont {K.}~\bibnamefont {Takahashi}}, \bibinfo {author} {\bibfnamefont {M.}~\bibnamefont {Kawasaki}},\ and\ \bibinfo {author} {\bibfnamefont {Y.}~\bibnamefont {Tokura}},\ }\bibfield  {title} {\bibinfo {title} {A magnetic heterostructure of topological insulators as a candidate for an axion insulator},\ }\href {https://doi.org/10.1038/nmat4855} {\bibfield  {journal} {\bibinfo  {journal} {Nat. Mater.}\ }\textbf {\bibinfo {volume} {16}},\ \bibinfo {pages} {516} (\bibinfo {year} {2017})}\BibitemShut {NoStop}%
	\bibitem [{\citenamefont {Ou}\ \emph {et~al.}(2018)\citenamefont {Ou}, \citenamefont {Liu}, \citenamefont {Jiang}, \citenamefont {Feng}, \citenamefont {Zhao}, \citenamefont {Wu}, \citenamefont {Wang}, \citenamefont {Li}, \citenamefont {Song}, \citenamefont {Wang}, \citenamefont {Wang}, \citenamefont {Wu}, \citenamefont {Wang}, \citenamefont {He}, \citenamefont {Ma},\ and\ \citenamefont {Xue}}]{ou2018enhancing}%
	\BibitemOpen
	\bibfield  {author} {\bibinfo {author} {\bibfnamefont {Y.}~\bibnamefont {Ou}}, \bibinfo {author} {\bibfnamefont {C.}~\bibnamefont {Liu}}, \bibinfo {author} {\bibfnamefont {G.}~\bibnamefont {Jiang}}, \bibinfo {author} {\bibfnamefont {Y.}~\bibnamefont {Feng}}, \bibinfo {author} {\bibfnamefont {D.}~\bibnamefont {Zhao}}, \bibinfo {author} {\bibfnamefont {W.}~\bibnamefont {Wu}}, \bibinfo {author} {\bibfnamefont {X.-X.}\ \bibnamefont {Wang}}, \bibinfo {author} {\bibfnamefont {W.}~\bibnamefont {Li}}, \bibinfo {author} {\bibfnamefont {C.}~\bibnamefont {Song}}, \bibinfo {author} {\bibfnamefont {L.-L.}\ \bibnamefont {Wang}}, \bibinfo {author} {\bibfnamefont {W.}~\bibnamefont {Wang}}, \bibinfo {author} {\bibfnamefont {W.}~\bibnamefont {Wu}}, \bibinfo {author} {\bibfnamefont {Y.}~\bibnamefont {Wang}}, \bibinfo {author} {\bibfnamefont {K.}~\bibnamefont {He}}, \bibinfo {author} {\bibfnamefont {X.-C.}\ \bibnamefont {Ma}},\ and\ \bibinfo {author} {\bibfnamefont {Q.-K.}\ \bibnamefont {Xue}},\ }\bibfield  {title} {\bibinfo
		{title} {Enhancing the quantum anomalous {Hall} effect by magnetic codoping in a topological insulator},\ }\href {https://doi.org/https://doi.org/10.1002/adma.201703062} {\bibfield  {journal} {\bibinfo  {journal} {Adv. Mater.}\ }\textbf {\bibinfo {volume} {30}},\ \bibinfo {pages} {1703062} (\bibinfo {year} {2018})}\BibitemShut {NoStop}%
	\bibitem [{\citenamefont {Deng}\ \emph {et~al.}(2020)\citenamefont {Deng}, \citenamefont {Yu}, \citenamefont {Shi}, \citenamefont {Guo}, \citenamefont {Xu}, \citenamefont {Wang}, \citenamefont {Chen},\ and\ \citenamefont {Zhang}}]{deng2020quantum}%
	\BibitemOpen
	\bibfield  {author} {\bibinfo {author} {\bibfnamefont {Y.}~\bibnamefont {Deng}}, \bibinfo {author} {\bibfnamefont {Y.}~\bibnamefont {Yu}}, \bibinfo {author} {\bibfnamefont {M.~Z.}\ \bibnamefont {Shi}}, \bibinfo {author} {\bibfnamefont {Z.}~\bibnamefont {Guo}}, \bibinfo {author} {\bibfnamefont {Z.}~\bibnamefont {Xu}}, \bibinfo {author} {\bibfnamefont {J.}~\bibnamefont {Wang}}, \bibinfo {author} {\bibfnamefont {X.~H.}\ \bibnamefont {Chen}},\ and\ \bibinfo {author} {\bibfnamefont {Y.}~\bibnamefont {Zhang}},\ }\bibfield  {title} {\bibinfo {title} {Quantum anomalous {Hall} effect in intrinsic magnetic topological insulator {MnBi$_2$Te$_4$}},\ }\href@noop {} {\bibfield  {journal} {\bibinfo  {journal} {Science}\ }\textbf {\bibinfo {volume} {367}},\ \bibinfo {pages} {895} (\bibinfo {year} {2020})}\BibitemShut {NoStop}%
	\bibitem [{\citenamefont {Serlin}\ \emph {et~al.}(2020)\citenamefont {Serlin}, \citenamefont {Tschirhart}, \citenamefont {Polshyn}, \citenamefont {Zhang}, \citenamefont {Zhu}, \citenamefont {Watanabe}, \citenamefont {Taniguchi}, \citenamefont {Balents},\ and\ \citenamefont {Young}}]{serlin2020intrinsic}%
	\BibitemOpen
	\bibfield  {author} {\bibinfo {author} {\bibfnamefont {M.}~\bibnamefont {Serlin}}, \bibinfo {author} {\bibfnamefont {C.~L.}\ \bibnamefont {Tschirhart}}, \bibinfo {author} {\bibfnamefont {H.}~\bibnamefont {Polshyn}}, \bibinfo {author} {\bibfnamefont {Y.}~\bibnamefont {Zhang}}, \bibinfo {author} {\bibfnamefont {J.}~\bibnamefont {Zhu}}, \bibinfo {author} {\bibfnamefont {K.}~\bibnamefont {Watanabe}}, \bibinfo {author} {\bibfnamefont {T.}~\bibnamefont {Taniguchi}}, \bibinfo {author} {\bibfnamefont {L.}~\bibnamefont {Balents}},\ and\ \bibinfo {author} {\bibfnamefont {A.~F.}\ \bibnamefont {Young}},\ }\bibfield  {title} {\bibinfo {title} {Intrinsic quantized anomalous {Hall} effect in a {moiré} heterostructure},\ }\href {https://doi.org/10.1126/science.aay5533} {\bibfield  {journal} {\bibinfo  {journal} {Science}\ }\textbf {\bibinfo {volume} {367}},\ \bibinfo {pages} {900} (\bibinfo {year} {2020})}\BibitemShut {NoStop}%
	\bibitem [{\citenamefont {Li}\ \emph {et~al.}(2021)\citenamefont {Li}, \citenamefont {Jiang}, \citenamefont {Shen}, \citenamefont {Zhang}, \citenamefont {Li}, \citenamefont {Tao}, \citenamefont {Devakul}, \citenamefont {Watanabe}, \citenamefont {Taniguchi}, \citenamefont {Fu}, \citenamefont {Shan},\ and\ \citenamefont {Mak}}]{li2021quantum}%
	\BibitemOpen
	\bibfield  {author} {\bibinfo {author} {\bibfnamefont {T.}~\bibnamefont {Li}}, \bibinfo {author} {\bibfnamefont {S.}~\bibnamefont {Jiang}}, \bibinfo {author} {\bibfnamefont {B.}~\bibnamefont {Shen}}, \bibinfo {author} {\bibfnamefont {Y.}~\bibnamefont {Zhang}}, \bibinfo {author} {\bibfnamefont {L.}~\bibnamefont {Li}}, \bibinfo {author} {\bibfnamefont {Z.}~\bibnamefont {Tao}}, \bibinfo {author} {\bibfnamefont {T.}~\bibnamefont {Devakul}}, \bibinfo {author} {\bibfnamefont {K.}~\bibnamefont {Watanabe}}, \bibinfo {author} {\bibfnamefont {T.}~\bibnamefont {Taniguchi}}, \bibinfo {author} {\bibfnamefont {L.}~\bibnamefont {Fu}}, \bibinfo {author} {\bibfnamefont {J.}~\bibnamefont {Shan}},\ and\ \bibinfo {author} {\bibfnamefont {K.~F.}\ \bibnamefont {Mak}},\ }\bibfield  {title} {\bibinfo {title} {Quantum anomalous {Hall} effect from intertwined moiré bands},\ }\href {https://doi.org/10.1038/s41586-021-04171-1} {\bibfield  {journal} {\bibinfo  {journal} {Nature}\ }\textbf {\bibinfo {volume} {600}},\ \bibinfo {pages}
		{641} (\bibinfo {year} {2021})}\BibitemShut {NoStop}%
	\bibitem [{\citenamefont {Wang}\ \emph {et~al.}(2013)\citenamefont {Wang}, \citenamefont {Lian}, \citenamefont {Zhang}, \citenamefont {Xu},\ and\ \citenamefont {Zhang}}]{wang2013quantum}%
	\BibitemOpen
	\bibfield  {author} {\bibinfo {author} {\bibfnamefont {J.}~\bibnamefont {Wang}}, \bibinfo {author} {\bibfnamefont {B.}~\bibnamefont {Lian}}, \bibinfo {author} {\bibfnamefont {H.}~\bibnamefont {Zhang}}, \bibinfo {author} {\bibfnamefont {Y.}~\bibnamefont {Xu}},\ and\ \bibinfo {author} {\bibfnamefont {S.-C.}\ \bibnamefont {Zhang}},\ }\bibfield  {title} {\bibinfo {title} {Quantum anomalous {Hall} effect with higher plateaus},\ }\href {https://doi.org/10.1103/PhysRevLett.111.136801} {\bibfield  {journal} {\bibinfo  {journal} {Phys. Rev. Lett.}\ }\textbf {\bibinfo {volume} {111}},\ \bibinfo {pages} {136801} (\bibinfo {year} {2013})}\BibitemShut {NoStop}%
	\bibitem [{\citenamefont {Du}\ \emph {et~al.}(2020)\citenamefont {Du}, \citenamefont {Tang}, \citenamefont {Li}, \citenamefont {Lin}, \citenamefont {Xu}, \citenamefont {Duan},\ and\ \citenamefont {Rubio}}]{du2020berry}%
	\BibitemOpen
	\bibfield  {author} {\bibinfo {author} {\bibfnamefont {S.}~\bibnamefont {Du}}, \bibinfo {author} {\bibfnamefont {P.}~\bibnamefont {Tang}}, \bibinfo {author} {\bibfnamefont {J.}~\bibnamefont {Li}}, \bibinfo {author} {\bibfnamefont {Z.}~\bibnamefont {Lin}}, \bibinfo {author} {\bibfnamefont {Y.}~\bibnamefont {Xu}}, \bibinfo {author} {\bibfnamefont {W.}~\bibnamefont {Duan}},\ and\ \bibinfo {author} {\bibfnamefont {A.}~\bibnamefont {Rubio}},\ }\bibfield  {title} {\bibinfo {title} {Berry curvature engineering by gating two-dimensional antiferromagnets},\ }\href {https://doi.org/10.1103/PhysRevResearch.2.022025} {\bibfield  {journal} {\bibinfo  {journal} {Phys. Rev. Res.}\ }\textbf {\bibinfo {volume} {2}},\ \bibinfo {pages} {022025} (\bibinfo {year} {2020})}\BibitemShut {NoStop}%
	\bibitem [{\citenamefont {Burkov}\ and\ \citenamefont {Balents}(2011)}]{burkov2011weyl}%
	\BibitemOpen
	\bibfield  {author} {\bibinfo {author} {\bibfnamefont {A.~A.}\ \bibnamefont {Burkov}}\ and\ \bibinfo {author} {\bibfnamefont {L.}~\bibnamefont {Balents}},\ }\bibfield  {title} {\bibinfo {title} {Weyl semimetal in a topological insulator multilayer},\ }\href {https://doi.org/10.1103/PhysRevLett.107.127205} {\bibfield  {journal} {\bibinfo  {journal} {Phys. Rev. Lett.}\ }\textbf {\bibinfo {volume} {107}},\ \bibinfo {pages} {127205} (\bibinfo {year} {2011})}\BibitemShut {NoStop}%
	\bibitem [{\citenamefont {Bosnar}\ \emph {et~al.}(2023)\citenamefont {Bosnar}, \citenamefont {Vyazovskaya}, \citenamefont {Petrov}, \citenamefont {Chulkov},\ and\ \citenamefont {Otrokov}}]{bosnar2023high}%
	\BibitemOpen
	\bibfield  {author} {\bibinfo {author} {\bibfnamefont {M.}~\bibnamefont {Bosnar}}, \bibinfo {author} {\bibfnamefont {A.~Y.}\ \bibnamefont {Vyazovskaya}}, \bibinfo {author} {\bibfnamefont {E.~K.}\ \bibnamefont {Petrov}}, \bibinfo {author} {\bibfnamefont {E.~V.}\ \bibnamefont {Chulkov}},\ and\ \bibinfo {author} {\bibfnamefont {M.~M.}\ \bibnamefont {Otrokov}},\ }\bibfield  {title} {\bibinfo {title} {High {Chern} number van der {Waals} magnetic topological multilayers {MnBi$_2$Te$_4$}/{hBN}},\ }\href {https://doi.org/10.1038/s41699-023-00396-y} {\bibfield  {journal} {\bibinfo  {journal} {npj 2D Mater. Appl.}\ }\textbf {\bibinfo {volume} {7}},\ \bibinfo {pages} {33} (\bibinfo {year} {2023})}\BibitemShut {NoStop}%
	\bibitem [{\citenamefont {Zhao}\ \emph {et~al.}(2020)\citenamefont {Zhao}, \citenamefont {Zhang}, \citenamefont {Mei}, \citenamefont {Zhou}, \citenamefont {Yi}, \citenamefont {Zhang}, \citenamefont {Yu}, \citenamefont {Xiao}, \citenamefont {Wang}, \citenamefont {Samarth}, \citenamefont {Chan}, \citenamefont {Liu},\ and\ \citenamefont {Chang}}]{zhao2020tuning}%
	\BibitemOpen
	\bibfield  {author} {\bibinfo {author} {\bibfnamefont {Y.-F.}\ \bibnamefont {Zhao}}, \bibinfo {author} {\bibfnamefont {R.}~\bibnamefont {Zhang}}, \bibinfo {author} {\bibfnamefont {R.}~\bibnamefont {Mei}}, \bibinfo {author} {\bibfnamefont {L.-J.}\ \bibnamefont {Zhou}}, \bibinfo {author} {\bibfnamefont {H.}~\bibnamefont {Yi}}, \bibinfo {author} {\bibfnamefont {Y.-Q.}\ \bibnamefont {Zhang}}, \bibinfo {author} {\bibfnamefont {J.}~\bibnamefont {Yu}}, \bibinfo {author} {\bibfnamefont {R.}~\bibnamefont {Xiao}}, \bibinfo {author} {\bibfnamefont {K.}~\bibnamefont {Wang}}, \bibinfo {author} {\bibfnamefont {N.}~\bibnamefont {Samarth}}, \bibinfo {author} {\bibfnamefont {M.~H.~W.}\ \bibnamefont {Chan}}, \bibinfo {author} {\bibfnamefont {C.-X.}\ \bibnamefont {Liu}},\ and\ \bibinfo {author} {\bibfnamefont {C.-Z.}\ \bibnamefont {Chang}},\ }\bibfield  {title} {\bibinfo {title} {Tuning the {Chern} number in quantum anomalous {Hall} insulators},\ }\href {https://doi.org/10.1038/s41586-020-3020-3} {\bibfield  {journal}
		{\bibinfo  {journal} {Nature}\ }\textbf {\bibinfo {volume} {588}},\ \bibinfo {pages} {419} (\bibinfo {year} {2020})}\BibitemShut {NoStop}%
	\bibitem [{\citenamefont {Chang}\ \emph {et~al.}(2023)\citenamefont {Chang}, \citenamefont {Liu},\ and\ \citenamefont {MacDonald}}]{chang2023quantum}%
	\BibitemOpen
	\bibfield  {author} {\bibinfo {author} {\bibfnamefont {C.-Z.}\ \bibnamefont {Chang}}, \bibinfo {author} {\bibfnamefont {C.-X.}\ \bibnamefont {Liu}},\ and\ \bibinfo {author} {\bibfnamefont {A.~H.}\ \bibnamefont {MacDonald}},\ }\bibfield  {title} {\bibinfo {title} {Colloquium: Quantum anomalous {Hall} effect},\ }\href {https://doi.org/10.1103/RevModPhys.95.011002} {\bibfield  {journal} {\bibinfo  {journal} {Rev. Mod. Phys.}\ }\textbf {\bibinfo {volume} {95}},\ \bibinfo {pages} {011002} (\bibinfo {year} {2023})}\BibitemShut {NoStop}%
	\bibitem [{\citenamefont {Li}\ \emph {et~al.}(2019{\natexlab{a}})\citenamefont {Li}, \citenamefont {Li}, \citenamefont {Du}, \citenamefont {Wang}, \citenamefont {Gu}, \citenamefont {Zhang}, \citenamefont {He}, \citenamefont {Duan},\ and\ \citenamefont {Xu}}]{li2019intrinsic}%
	\BibitemOpen
	\bibfield  {author} {\bibinfo {author} {\bibfnamefont {J.}~\bibnamefont {Li}}, \bibinfo {author} {\bibfnamefont {Y.}~\bibnamefont {Li}}, \bibinfo {author} {\bibfnamefont {S.}~\bibnamefont {Du}}, \bibinfo {author} {\bibfnamefont {Z.}~\bibnamefont {Wang}}, \bibinfo {author} {\bibfnamefont {B.-L.}\ \bibnamefont {Gu}}, \bibinfo {author} {\bibfnamefont {S.-C.}\ \bibnamefont {Zhang}}, \bibinfo {author} {\bibfnamefont {K.}~\bibnamefont {He}}, \bibinfo {author} {\bibfnamefont {W.}~\bibnamefont {Duan}},\ and\ \bibinfo {author} {\bibfnamefont {Y.}~\bibnamefont {Xu}},\ }\bibfield  {title} {\bibinfo {title} {Intrinsic magnetic topological insulators in van der {Waals} layered {MnBi$_2$Te$_4$}-family materials},\ }\href {https://doi.org/10.1126/sciadv.aaw5685} {\bibfield  {journal} {\bibinfo  {journal} {Sci. Adv.}\ }\textbf {\bibinfo {volume} {5}},\ \bibinfo {pages} {eaaw5685} (\bibinfo {year} {2019}{\natexlab{a}})}\BibitemShut {NoStop}%
	\bibitem [{\citenamefont {Zhang}\ \emph {et~al.}(2019)\citenamefont {Zhang}, \citenamefont {Shi}, \citenamefont {Zhu}, \citenamefont {Xing}, \citenamefont {Zhang},\ and\ \citenamefont {Wang}}]{zhang2019topological}%
	\BibitemOpen
	\bibfield  {author} {\bibinfo {author} {\bibfnamefont {D.}~\bibnamefont {Zhang}}, \bibinfo {author} {\bibfnamefont {M.}~\bibnamefont {Shi}}, \bibinfo {author} {\bibfnamefont {T.}~\bibnamefont {Zhu}}, \bibinfo {author} {\bibfnamefont {D.}~\bibnamefont {Xing}}, \bibinfo {author} {\bibfnamefont {H.}~\bibnamefont {Zhang}},\ and\ \bibinfo {author} {\bibfnamefont {J.}~\bibnamefont {Wang}},\ }\bibfield  {title} {\bibinfo {title} {Topological {Axion} states in the magnetic insulator {MnBi$_2$Te$_4$} with the quantized magnetoelectric effect},\ }\href {https://doi.org/10.1103/PhysRevLett.122.206401} {\bibfield  {journal} {\bibinfo  {journal} {Phys. Rev. Lett.}\ }\textbf {\bibinfo {volume} {122}},\ \bibinfo {pages} {206401} (\bibinfo {year} {2019})}\BibitemShut {NoStop}%
	\bibitem [{\citenamefont {Otrokov}\ \emph {et~al.}(2019{\natexlab{a}})\citenamefont {Otrokov}, \citenamefont {Rusinov}, \citenamefont {Blanco-Rey}, \citenamefont {Hoffmann}, \citenamefont {Vyazovskaya}, \citenamefont {Eremeev}, \citenamefont {Ernst}, \citenamefont {Echenique}, \citenamefont {Arnau},\ and\ \citenamefont {Chulkov}}]{otrokov2019unique}%
	\BibitemOpen
	\bibfield  {author} {\bibinfo {author} {\bibfnamefont {M.~M.}\ \bibnamefont {Otrokov}}, \bibinfo {author} {\bibfnamefont {I.~P.}\ \bibnamefont {Rusinov}}, \bibinfo {author} {\bibfnamefont {M.}~\bibnamefont {Blanco-Rey}}, \bibinfo {author} {\bibfnamefont {M.}~\bibnamefont {Hoffmann}}, \bibinfo {author} {\bibfnamefont {A.~Y.}\ \bibnamefont {Vyazovskaya}}, \bibinfo {author} {\bibfnamefont {S.~V.}\ \bibnamefont {Eremeev}}, \bibinfo {author} {\bibfnamefont {A.}~\bibnamefont {Ernst}}, \bibinfo {author} {\bibfnamefont {P.~M.}\ \bibnamefont {Echenique}}, \bibinfo {author} {\bibfnamefont {A.}~\bibnamefont {Arnau}},\ and\ \bibinfo {author} {\bibfnamefont {E.~V.}\ \bibnamefont {Chulkov}},\ }\bibfield  {title} {\bibinfo {title} {Unique thickness-dependent properties of the van der {Waals} interlayer antiferromagnet {${\mathrm{MnBi}}_{2}{\mathrm{Te}}_{4}$} films},\ }\href {https://doi.org/10.1103/PhysRevLett.122.107202} {\bibfield  {journal} {\bibinfo  {journal} {Phys. Rev. Lett.}\ }\textbf {\bibinfo {volume} {122}},\
		\bibinfo {pages} {107202} (\bibinfo {year} {2019}{\natexlab{a}})}\BibitemShut {NoStop}%
	\bibitem [{\citenamefont {Lee}\ \emph {et~al.}(2013)\citenamefont {Lee}, \citenamefont {Kim}, \citenamefont {Park}, \citenamefont {Chung}, \citenamefont {Lim}, \citenamefont {Seo},\ and\ \citenamefont {Park}}]{lee2013crystal}%
	\BibitemOpen
	\bibfield  {author} {\bibinfo {author} {\bibfnamefont {D.~S.}\ \bibnamefont {Lee}}, \bibinfo {author} {\bibfnamefont {T.-H.}\ \bibnamefont {Kim}}, \bibinfo {author} {\bibfnamefont {C.-H.}\ \bibnamefont {Park}}, \bibinfo {author} {\bibfnamefont {C.-Y.}\ \bibnamefont {Chung}}, \bibinfo {author} {\bibfnamefont {Y.~S.}\ \bibnamefont {Lim}}, \bibinfo {author} {\bibfnamefont {W.-S.}\ \bibnamefont {Seo}},\ and\ \bibinfo {author} {\bibfnamefont {H.-H.}\ \bibnamefont {Park}},\ }\bibfield  {title} {\bibinfo {title} {Crystal structure, properties and nanostructuring of a new layered chalcogenide semiconductor, {Bi$_2$MnTe$_4$}},\ }\href@noop {} {\bibfield  {journal} {\bibinfo  {journal} {Cryst. Eng. Comm.}\ }\textbf {\bibinfo {volume} {15}},\ \bibinfo {pages} {5532} (\bibinfo {year} {2013})}\BibitemShut {NoStop}%
	\bibitem [{\citenamefont {Gong}\ \emph {et~al.}(2019)\citenamefont {Gong}, \citenamefont {Guo}, \citenamefont {Li}, \citenamefont {Zhu}, \citenamefont {Liao}, \citenamefont {Liu}, \citenamefont {Zhang}, \citenamefont {Gu}, \citenamefont {Tang}, \citenamefont {Feng}, \citenamefont {Zhang}, \citenamefont {Li}, \citenamefont {Song}, \citenamefont {Wang}, \citenamefont {Yu}, \citenamefont {Chen}, \citenamefont {Wang}, \citenamefont {Yao}, \citenamefont {Duan}, \citenamefont {Xu}, \citenamefont {Zhang}, \citenamefont {Ma}, \citenamefont {Xue},\ and\ \citenamefont {He}}]{gong2019experimental}%
	\BibitemOpen
	\bibfield  {author} {\bibinfo {author} {\bibfnamefont {Y.}~\bibnamefont {Gong}}, \bibinfo {author} {\bibfnamefont {J.}~\bibnamefont {Guo}}, \bibinfo {author} {\bibfnamefont {J.}~\bibnamefont {Li}}, \bibinfo {author} {\bibfnamefont {K.}~\bibnamefont {Zhu}}, \bibinfo {author} {\bibfnamefont {M.}~\bibnamefont {Liao}}, \bibinfo {author} {\bibfnamefont {X.}~\bibnamefont {Liu}}, \bibinfo {author} {\bibfnamefont {Q.}~\bibnamefont {Zhang}}, \bibinfo {author} {\bibfnamefont {L.}~\bibnamefont {Gu}}, \bibinfo {author} {\bibfnamefont {L.}~\bibnamefont {Tang}}, \bibinfo {author} {\bibfnamefont {X.}~\bibnamefont {Feng}}, \bibinfo {author} {\bibfnamefont {D.}~\bibnamefont {Zhang}}, \bibinfo {author} {\bibfnamefont {W.}~\bibnamefont {Li}}, \bibinfo {author} {\bibfnamefont {C.}~\bibnamefont {Song}}, \bibinfo {author} {\bibfnamefont {L.}~\bibnamefont {Wang}}, \bibinfo {author} {\bibfnamefont {P.}~\bibnamefont {Yu}}, \bibinfo {author} {\bibfnamefont {X.}~\bibnamefont {Chen}}, \bibinfo {author} {\bibfnamefont {Y.}~\bibnamefont
			{Wang}}, \bibinfo {author} {\bibfnamefont {H.}~\bibnamefont {Yao}}, \bibinfo {author} {\bibfnamefont {W.}~\bibnamefont {Duan}}, \bibinfo {author} {\bibfnamefont {Y.}~\bibnamefont {Xu}}, \bibinfo {author} {\bibfnamefont {S.-C.}\ \bibnamefont {Zhang}}, \bibinfo {author} {\bibfnamefont {X.}~\bibnamefont {Ma}}, \bibinfo {author} {\bibfnamefont {Q.-K.}\ \bibnamefont {Xue}},\ and\ \bibinfo {author} {\bibfnamefont {K.}~\bibnamefont {He}},\ }\bibfield  {title} {\bibinfo {title} {Experimental realization of an intrinsic magnetic topological insulator*},\ }\href {https://doi.org/10.1088/0256-307X/36/7/076801} {\bibfield  {journal} {\bibinfo  {journal} {Chin. Phys. Lett.}\ }\textbf {\bibinfo {volume} {36}},\ \bibinfo {pages} {076801} (\bibinfo {year} {2019})}\BibitemShut {NoStop}%
	\bibitem [{\citenamefont {Yan}\ \emph {et~al.}(2019)\citenamefont {Yan}, \citenamefont {Zhang}, \citenamefont {Heitmann}, \citenamefont {Huang}, \citenamefont {Chen}, \citenamefont {Cheng}, \citenamefont {Wu}, \citenamefont {Vaknin}, \citenamefont {Sales},\ and\ \citenamefont {McQueeney}}]{yan2019crystal}%
	\BibitemOpen
	\bibfield  {author} {\bibinfo {author} {\bibfnamefont {J.-Q.}\ \bibnamefont {Yan}}, \bibinfo {author} {\bibfnamefont {Q.}~\bibnamefont {Zhang}}, \bibinfo {author} {\bibfnamefont {T.}~\bibnamefont {Heitmann}}, \bibinfo {author} {\bibfnamefont {Z.}~\bibnamefont {Huang}}, \bibinfo {author} {\bibfnamefont {K.~Y.}\ \bibnamefont {Chen}}, \bibinfo {author} {\bibfnamefont {J.-G.}\ \bibnamefont {Cheng}}, \bibinfo {author} {\bibfnamefont {W.}~\bibnamefont {Wu}}, \bibinfo {author} {\bibfnamefont {D.}~\bibnamefont {Vaknin}}, \bibinfo {author} {\bibfnamefont {B.~C.}\ \bibnamefont {Sales}},\ and\ \bibinfo {author} {\bibfnamefont {R.~J.}\ \bibnamefont {McQueeney}},\ }\bibfield  {title} {\bibinfo {title} {Crystal growth and magnetic structure of {${\mathrm{MnBi}}_{2}{\mathrm{Te}}_{4}$}},\ }\href {https://doi.org/10.1103/PhysRevMaterials.3.064202} {\bibfield  {journal} {\bibinfo  {journal} {Phys. Rev. Mater.}\ }\textbf {\bibinfo {volume} {3}},\ \bibinfo {pages} {064202} (\bibinfo {year} {2019})}\BibitemShut {NoStop}%
	\bibitem [{\citenamefont {Cui}\ \emph {et~al.}(2019)\citenamefont {Cui}, \citenamefont {Shi}, \citenamefont {Wang}, \citenamefont {Yu}, \citenamefont {Wu}, \citenamefont {Luo}, \citenamefont {Ying},\ and\ \citenamefont {Chen}}]{cui2019transport}%
	\BibitemOpen
	\bibfield  {author} {\bibinfo {author} {\bibfnamefont {J.}~\bibnamefont {Cui}}, \bibinfo {author} {\bibfnamefont {M.}~\bibnamefont {Shi}}, \bibinfo {author} {\bibfnamefont {H.}~\bibnamefont {Wang}}, \bibinfo {author} {\bibfnamefont {F.}~\bibnamefont {Yu}}, \bibinfo {author} {\bibfnamefont {T.}~\bibnamefont {Wu}}, \bibinfo {author} {\bibfnamefont {X.}~\bibnamefont {Luo}}, \bibinfo {author} {\bibfnamefont {J.}~\bibnamefont {Ying}},\ and\ \bibinfo {author} {\bibfnamefont {X.}~\bibnamefont {Chen}},\ }\bibfield  {title} {\bibinfo {title} {Transport properties of thin flakes of the antiferromagnetic topological insulator $\mathrm{MnB}{\mathrm{i}}_{2}\mathrm{T}{\mathrm{e}}_{4}$},\ }\href {https://doi.org/10.1103/PhysRevB.99.155125} {\bibfield  {journal} {\bibinfo  {journal} {Phys. Rev. B}\ }\textbf {\bibinfo {volume} {99}},\ \bibinfo {pages} {155125} (\bibinfo {year} {2019})}\BibitemShut {NoStop}%
	\bibitem [{\citenamefont {Li}\ \emph {et~al.}(2019{\natexlab{b}})\citenamefont {Li}, \citenamefont {Wang}, \citenamefont {Zhang}, \citenamefont {Gu}, \citenamefont {Duan},\ and\ \citenamefont {Xu}}]{li2019magnetically}%
	\BibitemOpen
	\bibfield  {author} {\bibinfo {author} {\bibfnamefont {J.}~\bibnamefont {Li}}, \bibinfo {author} {\bibfnamefont {C.}~\bibnamefont {Wang}}, \bibinfo {author} {\bibfnamefont {Z.}~\bibnamefont {Zhang}}, \bibinfo {author} {\bibfnamefont {B.-L.}\ \bibnamefont {Gu}}, \bibinfo {author} {\bibfnamefont {W.}~\bibnamefont {Duan}},\ and\ \bibinfo {author} {\bibfnamefont {Y.}~\bibnamefont {Xu}},\ }\bibfield  {title} {\bibinfo {title} {Magnetically controllable topological quantum phase transitions in the antiferromagnetic topological insulator {MnBi$_2$Te$_4$}},\ }\href {https://doi.org/10.1103/PhysRevB.100.121103} {\bibfield  {journal} {\bibinfo  {journal} {Phys. Rev. B}\ }\textbf {\bibinfo {volume} {100}},\ \bibinfo {pages} {121103} (\bibinfo {year} {2019}{\natexlab{b}})}\BibitemShut {NoStop}%
	\bibitem [{\citenamefont {Liu}\ \emph {et~al.}(2020)\citenamefont {Liu}, \citenamefont {Wang}, \citenamefont {Li}, \citenamefont {Wu}, \citenamefont {Li}, \citenamefont {Li}, \citenamefont {He}, \citenamefont {Xu}, \citenamefont {Zhang},\ and\ \citenamefont {Wang}}]{liu2020robust}%
	\BibitemOpen
	\bibfield  {author} {\bibinfo {author} {\bibfnamefont {C.}~\bibnamefont {Liu}}, \bibinfo {author} {\bibfnamefont {Y.}~\bibnamefont {Wang}}, \bibinfo {author} {\bibfnamefont {H.}~\bibnamefont {Li}}, \bibinfo {author} {\bibfnamefont {Y.}~\bibnamefont {Wu}}, \bibinfo {author} {\bibfnamefont {Y.}~\bibnamefont {Li}}, \bibinfo {author} {\bibfnamefont {J.}~\bibnamefont {Li}}, \bibinfo {author} {\bibfnamefont {K.}~\bibnamefont {He}}, \bibinfo {author} {\bibfnamefont {Y.}~\bibnamefont {Xu}}, \bibinfo {author} {\bibfnamefont {J.}~\bibnamefont {Zhang}},\ and\ \bibinfo {author} {\bibfnamefont {Y.}~\bibnamefont {Wang}},\ }\bibfield  {title} {\bibinfo {title} {Robust axion insulator and {Chern} insulator phases in a two-dimensional antiferromagnetic topological insulator},\ }\href {https://doi.org/10.1038/s41563-019-0573-3} {\bibfield  {journal} {\bibinfo  {journal} {Nat. Mater.}\ }\textbf {\bibinfo {volume} {19}},\ \bibinfo {pages} {522} (\bibinfo {year} {2020})}\BibitemShut {NoStop}%
	\bibitem [{\citenamefont {Chen}\ \emph {et~al.}(2019{\natexlab{a}})\citenamefont {Chen}, \citenamefont {Xu}, \citenamefont {Li}, \citenamefont {Li}, \citenamefont {Wang}, \citenamefont {Zhang}, \citenamefont {Li}, \citenamefont {Wu}, \citenamefont {Liang}, \citenamefont {Chen}, \citenamefont {Jung}, \citenamefont {Cacho}, \citenamefont {Mao}, \citenamefont {Liu}, \citenamefont {Wang}, \citenamefont {Guo}, \citenamefont {Xu}, \citenamefont {Liu}, \citenamefont {Yang},\ and\ \citenamefont {Chen}}]{chen2019topological}%
	\BibitemOpen
	\bibfield  {author} {\bibinfo {author} {\bibfnamefont {Y.~J.}\ \bibnamefont {Chen}}, \bibinfo {author} {\bibfnamefont {L.~X.}\ \bibnamefont {Xu}}, \bibinfo {author} {\bibfnamefont {J.~H.}\ \bibnamefont {Li}}, \bibinfo {author} {\bibfnamefont {Y.~W.}\ \bibnamefont {Li}}, \bibinfo {author} {\bibfnamefont {H.~Y.}\ \bibnamefont {Wang}}, \bibinfo {author} {\bibfnamefont {C.~F.}\ \bibnamefont {Zhang}}, \bibinfo {author} {\bibfnamefont {H.}~\bibnamefont {Li}}, \bibinfo {author} {\bibfnamefont {Y.}~\bibnamefont {Wu}}, \bibinfo {author} {\bibfnamefont {A.~J.}\ \bibnamefont {Liang}}, \bibinfo {author} {\bibfnamefont {C.}~\bibnamefont {Chen}}, \bibinfo {author} {\bibfnamefont {S.~W.}\ \bibnamefont {Jung}}, \bibinfo {author} {\bibfnamefont {C.}~\bibnamefont {Cacho}}, \bibinfo {author} {\bibfnamefont {Y.~H.}\ \bibnamefont {Mao}}, \bibinfo {author} {\bibfnamefont {S.}~\bibnamefont {Liu}}, \bibinfo {author} {\bibfnamefont {M.~X.}\ \bibnamefont {Wang}}, \bibinfo {author} {\bibfnamefont {Y.~F.}\ \bibnamefont {Guo}},
		\bibinfo {author} {\bibfnamefont {Y.}~\bibnamefont {Xu}}, \bibinfo {author} {\bibfnamefont {Z.~K.}\ \bibnamefont {Liu}}, \bibinfo {author} {\bibfnamefont {L.~X.}\ \bibnamefont {Yang}},\ and\ \bibinfo {author} {\bibfnamefont {Y.~L.}\ \bibnamefont {Chen}},\ }\bibfield  {title} {\bibinfo {title} {Topological electronic structure and its temperature evolution in antiferromagnetic topological insulator {MnBi$_2$Te$_4$}},\ }\href {https://doi.org/10.1103/PhysRevX.9.041040} {\bibfield  {journal} {\bibinfo  {journal} {Phys. Rev. X}\ }\textbf {\bibinfo {volume} {9}},\ \bibinfo {pages} {041040} (\bibinfo {year} {2019}{\natexlab{a}})}\BibitemShut {NoStop}%
	\bibitem [{\citenamefont {Otrokov}\ \emph {et~al.}(2019{\natexlab{b}})\citenamefont {Otrokov}, \citenamefont {Klimovskikh}, \citenamefont {Bentmann}, \citenamefont {Estyunin}, \citenamefont {Zeugner}, \citenamefont {Aliev}, \citenamefont {Gaß}, \citenamefont {Wolter}, \citenamefont {Koroleva}, \citenamefont {Shikin}, \citenamefont {Blanco-Rey}, \citenamefont {Hoffmann}, \citenamefont {Rusinov}, \citenamefont {Vyazovskaya}, \citenamefont {Eremeev}, \citenamefont {Koroteev}, \citenamefont {Kuznetsov}, \citenamefont {Freyse}, \citenamefont {Sánchez-Barriga}, \citenamefont {Amiraslanov}, \citenamefont {Babanly}, \citenamefont {Mamedov}, \citenamefont {Abdullayev}, \citenamefont {Zverev}, \citenamefont {Alfonsov}, \citenamefont {Kataev}, \citenamefont {Büchner}, \citenamefont {Schwier}, \citenamefont {Kumar}, \citenamefont {Kimura}, \citenamefont {Petaccia}, \citenamefont {Di~Santo}, \citenamefont {Vidal}, \citenamefont {Schatz}, \citenamefont {Kißner}, \citenamefont {Ünzelmann}, \citenamefont {Min},
		\citenamefont {Moser}, \citenamefont {Peixoto}, \citenamefont {Reinert}, \citenamefont {Ernst}, \citenamefont {Echenique}, \citenamefont {Isaeva},\ and\ \citenamefont {Chulkov}}]{otrokov2019prediction}%
	\BibitemOpen
	\bibfield  {author} {\bibinfo {author} {\bibfnamefont {M.~M.}\ \bibnamefont {Otrokov}}, \bibinfo {author} {\bibfnamefont {I.~I.}\ \bibnamefont {Klimovskikh}}, \bibinfo {author} {\bibfnamefont {H.}~\bibnamefont {Bentmann}}, \bibinfo {author} {\bibfnamefont {D.}~\bibnamefont {Estyunin}}, \bibinfo {author} {\bibfnamefont {A.}~\bibnamefont {Zeugner}}, \bibinfo {author} {\bibfnamefont {Z.~S.}\ \bibnamefont {Aliev}}, \bibinfo {author} {\bibfnamefont {S.}~\bibnamefont {Gaß}}, \bibinfo {author} {\bibfnamefont {A.~U.~B.}\ \bibnamefont {Wolter}}, \bibinfo {author} {\bibfnamefont {A.~V.}\ \bibnamefont {Koroleva}}, \bibinfo {author} {\bibfnamefont {A.~M.}\ \bibnamefont {Shikin}}, \bibinfo {author} {\bibfnamefont {M.}~\bibnamefont {Blanco-Rey}}, \bibinfo {author} {\bibfnamefont {M.}~\bibnamefont {Hoffmann}}, \bibinfo {author} {\bibfnamefont {I.~P.}\ \bibnamefont {Rusinov}}, \bibinfo {author} {\bibfnamefont {A.~Y.}\ \bibnamefont {Vyazovskaya}}, \bibinfo {author} {\bibfnamefont {S.~V.}\ \bibnamefont {Eremeev}}, \bibinfo
		{author} {\bibfnamefont {Y.~M.}\ \bibnamefont {Koroteev}}, \bibinfo {author} {\bibfnamefont {V.~M.}\ \bibnamefont {Kuznetsov}}, \bibinfo {author} {\bibfnamefont {F.}~\bibnamefont {Freyse}}, \bibinfo {author} {\bibfnamefont {J.}~\bibnamefont {Sánchez-Barriga}}, \bibinfo {author} {\bibfnamefont {I.~R.}\ \bibnamefont {Amiraslanov}}, \bibinfo {author} {\bibfnamefont {M.~B.}\ \bibnamefont {Babanly}}, \bibinfo {author} {\bibfnamefont {N.~T.}\ \bibnamefont {Mamedov}}, \bibinfo {author} {\bibfnamefont {N.~A.}\ \bibnamefont {Abdullayev}}, \bibinfo {author} {\bibfnamefont {V.~N.}\ \bibnamefont {Zverev}}, \bibinfo {author} {\bibfnamefont {A.}~\bibnamefont {Alfonsov}}, \bibinfo {author} {\bibfnamefont {V.}~\bibnamefont {Kataev}}, \bibinfo {author} {\bibfnamefont {B.}~\bibnamefont {Büchner}}, \bibinfo {author} {\bibfnamefont {E.~F.}\ \bibnamefont {Schwier}}, \bibinfo {author} {\bibfnamefont {S.}~\bibnamefont {Kumar}}, \bibinfo {author} {\bibfnamefont {A.}~\bibnamefont {Kimura}}, \bibinfo {author} {\bibfnamefont
			{L.}~\bibnamefont {Petaccia}}, \bibinfo {author} {\bibfnamefont {G.}~\bibnamefont {Di~Santo}}, \bibinfo {author} {\bibfnamefont {R.~C.}\ \bibnamefont {Vidal}}, \bibinfo {author} {\bibfnamefont {S.}~\bibnamefont {Schatz}}, \bibinfo {author} {\bibfnamefont {K.}~\bibnamefont {Kißner}}, \bibinfo {author} {\bibfnamefont {M.}~\bibnamefont {Ünzelmann}}, \bibinfo {author} {\bibfnamefont {C.~H.}\ \bibnamefont {Min}}, \bibinfo {author} {\bibfnamefont {S.}~\bibnamefont {Moser}}, \bibinfo {author} {\bibfnamefont {T.~R.~F.}\ \bibnamefont {Peixoto}}, \bibinfo {author} {\bibfnamefont {F.}~\bibnamefont {Reinert}}, \bibinfo {author} {\bibfnamefont {A.}~\bibnamefont {Ernst}}, \bibinfo {author} {\bibfnamefont {P.~M.}\ \bibnamefont {Echenique}}, \bibinfo {author} {\bibfnamefont {A.}~\bibnamefont {Isaeva}},\ and\ \bibinfo {author} {\bibfnamefont {E.~V.}\ \bibnamefont {Chulkov}},\ }\bibfield  {title} {\bibinfo {title} {Prediction and observation of an antiferromagnetic topological insulator},\ }\href
	{https://doi.org/10.1038/s41586-019-1840-9} {\bibfield  {journal} {\bibinfo  {journal} {Nature}\ }\textbf {\bibinfo {volume} {576}},\ \bibinfo {pages} {416} (\bibinfo {year} {2019}{\natexlab{b}})}\BibitemShut {NoStop}%
	\bibitem [{\citenamefont {Zhu}\ \emph {et~al.}(2021)\citenamefont {Zhu}, \citenamefont {Wang}, \citenamefont {Zhang},\ and\ \citenamefont {Xing}}]{zhu2021tunable}%
	\BibitemOpen
	\bibfield  {author} {\bibinfo {author} {\bibfnamefont {T.}~\bibnamefont {Zhu}}, \bibinfo {author} {\bibfnamefont {H.}~\bibnamefont {Wang}}, \bibinfo {author} {\bibfnamefont {H.}~\bibnamefont {Zhang}},\ and\ \bibinfo {author} {\bibfnamefont {D.}~\bibnamefont {Xing}},\ }\bibfield  {title} {\bibinfo {title} {Tunable dynamical magnetoelectric effect in antiferromagnetic topological insulator {MnBi$_2$Te$_4$} films},\ }\href {https://doi.org/10.1038/s41524-021-00589-3} {\bibfield  {journal} {\bibinfo  {journal} {npj Comput. Mater.}\ }\textbf {\bibinfo {volume} {7}},\ \bibinfo {pages} {121} (\bibinfo {year} {2021})}\BibitemShut {NoStop}%
	\bibitem [{\citenamefont {Ge}\ \emph {et~al.}(2020)\citenamefont {Ge}, \citenamefont {Liu}, \citenamefont {Li}, \citenamefont {Li}, \citenamefont {Luo}, \citenamefont {Wu}, \citenamefont {Xu},\ and\ \citenamefont {Wang}}]{ge2020high}%
	\BibitemOpen
	\bibfield  {author} {\bibinfo {author} {\bibfnamefont {J.}~\bibnamefont {Ge}}, \bibinfo {author} {\bibfnamefont {Y.}~\bibnamefont {Liu}}, \bibinfo {author} {\bibfnamefont {J.}~\bibnamefont {Li}}, \bibinfo {author} {\bibfnamefont {H.}~\bibnamefont {Li}}, \bibinfo {author} {\bibfnamefont {T.}~\bibnamefont {Luo}}, \bibinfo {author} {\bibfnamefont {Y.}~\bibnamefont {Wu}}, \bibinfo {author} {\bibfnamefont {Y.}~\bibnamefont {Xu}},\ and\ \bibinfo {author} {\bibfnamefont {J.}~\bibnamefont {Wang}},\ }\bibfield  {title} {\bibinfo {title} {{High-Chern-number and high-temperature quantum Hall effect without Landau levels}},\ }\href {https://doi.org/10.1093/nsr/nwaa089} {\bibfield  {journal} {\bibinfo  {journal} {Natl. Sci. Rev.}\ }\textbf {\bibinfo {volume} {7}},\ \bibinfo {pages} {1280} (\bibinfo {year} {2020})}\BibitemShut {NoStop}%
	\bibitem [{\citenamefont {Ren}\ \emph {et~al.}(2022)\citenamefont {Ren}, \citenamefont {Ke}, \citenamefont {Lou},\ and\ \citenamefont {Chang}}]{ren2022quantum}%
	\BibitemOpen
	\bibfield  {author} {\bibinfo {author} {\bibfnamefont {Y.}~\bibnamefont {Ren}}, \bibinfo {author} {\bibfnamefont {S.}~\bibnamefont {Ke}}, \bibinfo {author} {\bibfnamefont {W.-K.}\ \bibnamefont {Lou}},\ and\ \bibinfo {author} {\bibfnamefont {K.}~\bibnamefont {Chang}},\ }\bibfield  {title} {\bibinfo {title} {Quantum phase transitions driven by sliding in bilayer {MnBi$_2$Te$_4$}},\ }\href {https://doi.org/10.1103/PhysRevB.106.235302} {\bibfield  {journal} {\bibinfo  {journal} {Phys. Rev. B}\ }\textbf {\bibinfo {volume} {106}},\ \bibinfo {pages} {235302} (\bibinfo {year} {2022})}\BibitemShut {NoStop}%
	\bibitem [{\citenamefont {Zhu}\ \emph {et~al.}(2022)\citenamefont {Zhu}, \citenamefont {Song}, \citenamefont {Bai}, \citenamefont {Liao},\ and\ \citenamefont {Pan}}]{zhu2022high}%
	\BibitemOpen
	\bibfield  {author} {\bibinfo {author} {\bibfnamefont {W.}~\bibnamefont {Zhu}}, \bibinfo {author} {\bibfnamefont {C.}~\bibnamefont {Song}}, \bibinfo {author} {\bibfnamefont {H.}~\bibnamefont {Bai}}, \bibinfo {author} {\bibfnamefont {L.}~\bibnamefont {Liao}},\ and\ \bibinfo {author} {\bibfnamefont {F.}~\bibnamefont {Pan}},\ }\bibfield  {title} {\bibinfo {title} {High {Chern} number quantum anomalous {Hall} effect tunable by stacking order in van der {Waals} topological insulators},\ }\href {https://doi.org/10.1103/PhysRevB.105.155122} {\bibfield  {journal} {\bibinfo  {journal} {Phys. Rev. B}\ }\textbf {\bibinfo {volume} {105}},\ \bibinfo {pages} {155122} (\bibinfo {year} {2022})}\BibitemShut {NoStop}%
	\bibitem [{\citenamefont {Cao}\ \emph {et~al.}(2023)\citenamefont {Cao}, \citenamefont {Shao}, \citenamefont {Huang}, \citenamefont {Gurung},\ and\ \citenamefont {Tsymbal}}]{cao2023switchable}%
	\BibitemOpen
	\bibfield  {author} {\bibinfo {author} {\bibfnamefont {T.}~\bibnamefont {Cao}}, \bibinfo {author} {\bibfnamefont {D.-F.}\ \bibnamefont {Shao}}, \bibinfo {author} {\bibfnamefont {K.}~\bibnamefont {Huang}}, \bibinfo {author} {\bibfnamefont {G.}~\bibnamefont {Gurung}},\ and\ \bibinfo {author} {\bibfnamefont {E.~Y.}\ \bibnamefont {Tsymbal}},\ }\bibfield  {title} {\bibinfo {title} {Switchable anomalous {Hall} effects in polar-stacked {2D} antiferromagnet {MnBi$_2$Te$_4$}},\ }\href@noop {} {\bibfield  {journal} {\bibinfo  {journal} {Nano Lett.}\ }\textbf {\bibinfo {volume} {23}},\ \bibinfo {pages} {3781} (\bibinfo {year} {2023})}\BibitemShut {NoStop}%
	\bibitem [{\citenamefont {Ahn}\ \emph {et~al.}(2023)\citenamefont {Ahn}, \citenamefont {Kang}, \citenamefont {Yoon}, \citenamefont {Ganesh},\ and\ \citenamefont {Krogel}}]{ahn2023stacking}%
	\BibitemOpen
	\bibfield  {author} {\bibinfo {author} {\bibfnamefont {J.}~\bibnamefont {Ahn}}, \bibinfo {author} {\bibfnamefont {S.-H.}\ \bibnamefont {Kang}}, \bibinfo {author} {\bibfnamefont {M.}~\bibnamefont {Yoon}}, \bibinfo {author} {\bibfnamefont {P.}~\bibnamefont {Ganesh}},\ and\ \bibinfo {author} {\bibfnamefont {J.~T.}\ \bibnamefont {Krogel}},\ }\bibfield  {title} {\bibinfo {title} {Stacking faults and topological properties in {MnBi$_2$Te$_4$}: Reconciling gapped and gapless states},\ }\href {https://doi.org/10.1021/acs.jpclett.3c01939} {\bibfield  {journal} {\bibinfo  {journal} {J. Phys. Chem. Lett.}\ }\textbf {\bibinfo {volume} {14}},\ \bibinfo {pages} {9052} (\bibinfo {year} {2023})}\BibitemShut {NoStop}%
	\bibitem [{\citenamefont {Li}\ \emph {et~al.}(2025)\citenamefont {Li}, \citenamefont {Wu},\ and\ \citenamefont {Weng}}]{li2025stacking}%
	\BibitemOpen
	\bibfield  {author} {\bibinfo {author} {\bibfnamefont {J.}~\bibnamefont {Li}}, \bibinfo {author} {\bibfnamefont {Q.}~\bibnamefont {Wu}},\ and\ \bibinfo {author} {\bibfnamefont {H.}~\bibnamefont {Weng}},\ }\bibfield  {title} {\bibinfo {title} {Stacking-dependent electronic and topological properties in van der {Waals} antiferromagnet {MnBi$_2$Te$_4$} films},\ }\href {https://doi.org/10.1038/s41524-025-01545-1} {\bibfield  {journal} {\bibinfo  {journal} {npj Comput. Mater.}\ }\textbf {\bibinfo {volume} {11}},\ \bibinfo {pages} {53} (\bibinfo {year} {2025})}\BibitemShut {NoStop}%
	\bibitem [{\citenamefont {Wang}\ \emph {et~al.}(2023)\citenamefont {Wang}, \citenamefont {Wang}, \citenamefont {Xing},\ and\ \citenamefont {Zhang}}]{wang2023three}%
	\BibitemOpen
	\bibfield  {author} {\bibinfo {author} {\bibfnamefont {D.}~\bibnamefont {Wang}}, \bibinfo {author} {\bibfnamefont {H.}~\bibnamefont {Wang}}, \bibinfo {author} {\bibfnamefont {D.}~\bibnamefont {Xing}},\ and\ \bibinfo {author} {\bibfnamefont {H.}~\bibnamefont {Zhang}},\ }\bibfield  {title} {\bibinfo {title} {Three-{Dirac}-fermion approach to unexpected universal gapless surface states in van der {Waals} magnetic topological insulators},\ }\href {https://doi.org/10.1007/s11433-023-2161-9} {\bibfield  {journal} {\bibinfo  {journal} {Sci. China Phys Mech.}\ }\textbf {\bibinfo {volume} {66}},\ \bibinfo {pages} {297211} (\bibinfo {year} {2023})}\BibitemShut {NoStop}%
	\bibitem [{SM()}]{SM}%
	\BibitemOpen
	\href@noop {} {}\bibinfo {note} {See Supplemental Material for detailed calculation methods, tight-binding parameters of Hamiltonians, and electronic band structures and real-space anomalous Hall conductivity of other typical $(X, -Y, X)$ and $(X, -Y, X, -Y, X)$ thin films with mixed stacking order.}\BibitemShut {Stop}%
	\bibitem [{\citenamefont {Bianco}\ and\ \citenamefont {Resta}(2011)}]{bianco2011mapping}%
	\BibitemOpen
	\bibfield  {author} {\bibinfo {author} {\bibfnamefont {R.}~\bibnamefont {Bianco}}\ and\ \bibinfo {author} {\bibfnamefont {R.}~\bibnamefont {Resta}},\ }\bibfield  {title} {\bibinfo {title} {Mapping topological order in coordinate space},\ }\href {https://doi.org/10.1103/PhysRevB.84.241106} {\bibfield  {journal} {\bibinfo  {journal} {Phys. Rev. B}\ }\textbf {\bibinfo {volume} {84}},\ \bibinfo {pages} {241106} (\bibinfo {year} {2011})}\BibitemShut {NoStop}%
	\bibitem [{\citenamefont {Marrazzo}\ and\ \citenamefont {Resta}(2017)}]{marrazzo2017locality}%
	\BibitemOpen
	\bibfield  {author} {\bibinfo {author} {\bibfnamefont {A.}~\bibnamefont {Marrazzo}}\ and\ \bibinfo {author} {\bibfnamefont {R.}~\bibnamefont {Resta}},\ }\bibfield  {title} {\bibinfo {title} {Locality of the anomalous {Hall} conductivity},\ }\href {https://doi.org/10.1103/PhysRevB.95.121114} {\bibfield  {journal} {\bibinfo  {journal} {Phys. Rev. B}\ }\textbf {\bibinfo {volume} {95}},\ \bibinfo {pages} {121114} (\bibinfo {year} {2017})}\BibitemShut {NoStop}%
	\bibitem [{\citenamefont {Varnava}\ and\ \citenamefont {Vanderbilt}(2018)}]{varnava2018surface}%
	\BibitemOpen
	\bibfield  {author} {\bibinfo {author} {\bibfnamefont {N.}~\bibnamefont {Varnava}}\ and\ \bibinfo {author} {\bibfnamefont {D.}~\bibnamefont {Vanderbilt}},\ }\bibfield  {title} {\bibinfo {title} {Surfaces of axion insulators},\ }\href {https://doi.org/10.1103/PhysRevB.98.245117} {\bibfield  {journal} {\bibinfo  {journal} {Phys. Rev. B}\ }\textbf {\bibinfo {volume} {98}},\ \bibinfo {pages} {245117} (\bibinfo {year} {2018})}\BibitemShut {NoStop}%
	\bibitem [{\citenamefont {Bernevig}\ \emph {et~al.}(2006)\citenamefont {Bernevig}, \citenamefont {Hughes},\ and\ \citenamefont {Zhang}}]{bernevig2006quantum}%
	\BibitemOpen
	\bibfield  {author} {\bibinfo {author} {\bibfnamefont {B.~A.}\ \bibnamefont {Bernevig}}, \bibinfo {author} {\bibfnamefont {T.~L.}\ \bibnamefont {Hughes}},\ and\ \bibinfo {author} {\bibfnamefont {S.-C.}\ \bibnamefont {Zhang}},\ }\bibfield  {title} {\bibinfo {title} {Quantum spin {Hall} effect and topological phase transition in {HgTe} quantum wells},\ }\href {https://doi.org/10.1126/science.1133734} {\bibfield  {journal} {\bibinfo  {journal} {Science}\ }\textbf {\bibinfo {volume} {314}},\ \bibinfo {pages} {1757} (\bibinfo {year} {2006})}\BibitemShut {NoStop}%
	\bibitem [{\citenamefont {Zhang}\ \emph {et~al.}(2009)\citenamefont {Zhang}, \citenamefont {Liu}, \citenamefont {Qi}, \citenamefont {Dai}, \citenamefont {Fang},\ and\ \citenamefont {Zhang}}]{zhang2009topological}%
	\BibitemOpen
	\bibfield  {author} {\bibinfo {author} {\bibfnamefont {H.}~\bibnamefont {Zhang}}, \bibinfo {author} {\bibfnamefont {C.-X.}\ \bibnamefont {Liu}}, \bibinfo {author} {\bibfnamefont {X.-L.}\ \bibnamefont {Qi}}, \bibinfo {author} {\bibfnamefont {X.}~\bibnamefont {Dai}}, \bibinfo {author} {\bibfnamefont {Z.}~\bibnamefont {Fang}},\ and\ \bibinfo {author} {\bibfnamefont {S.-C.}\ \bibnamefont {Zhang}},\ }\bibfield  {title} {\bibinfo {title} {Topological insulators in {Bi$_2$Se$_3$}, {Bi$_2$Te$_3$} and {Sb$_2$Te$_3$} with a single {Dirac} cone on the surface},\ }\href {https://doi.org/10.1038/nphys1270} {\bibfield  {journal} {\bibinfo  {journal} {Nat. Phys.}\ }\textbf {\bibinfo {volume} {5}},\ \bibinfo {pages} {438} (\bibinfo {year} {2009})}\BibitemShut {NoStop}%
	\bibitem [{\citenamefont {Liu}\ \emph {et~al.}(2010)\citenamefont {Liu}, \citenamefont {Qi}, \citenamefont {Zhang}, \citenamefont {Dai}, \citenamefont {Fang},\ and\ \citenamefont {Zhang}}]{liu2010model}
	\BibitemOpen
	\bibfield  {author} {\bibinfo {author} {\bibfnamefont {C.-X.}\ \bibnamefont {Liu}}, \bibinfo {author} {\bibfnamefont {X.-L.}\ \bibnamefont {Qi}}, \bibinfo {author} {\bibfnamefont {H.}~\bibnamefont {Zhang}}, \bibinfo {author} {\bibfnamefont {X.}~\bibnamefont {Dai}}, \bibinfo {author} {\bibfnamefont {Z.}~\bibnamefont {Fang}},\ and\ \bibinfo {author} {\bibfnamefont {S.-C.}\ \bibnamefont {Zhang}},\ }\bibfield  {title} {\bibinfo {title} {Model hamiltonian for topological insulators},\ }\href {https://doi.org/10.1103/PhysRevB.82.045122} {\bibfield  {journal} {\bibinfo  {journal} {Phys. Rev. B}\ }\textbf {\bibinfo {volume} {82}},\ \bibinfo {pages} {045122} (\bibinfo {year} {2010})}\BibitemShut {NoStop}%
	\bibitem [{\citenamefont {Li}\ \emph {et~al.}(2019{\natexlab{c}})\citenamefont {Li}, \citenamefont {Gao}, \citenamefont {Duan}, \citenamefont {Xu}, \citenamefont {Zhu}, \citenamefont {Tian}, \citenamefont {Gao}, \citenamefont {Fan}, \citenamefont {Rao}, \citenamefont {Huang}, \citenamefont {Li}, \citenamefont {Yan}, \citenamefont {Liu}, \citenamefont {Liu}, \citenamefont {Huang}, \citenamefont {Li}, \citenamefont {Liu}, \citenamefont {Zhang}, \citenamefont {Zhang}, \citenamefont {Kondo}, \citenamefont {Shin}, \citenamefont {Lei}, \citenamefont {Shi}, \citenamefont {Zhang}, \citenamefont {Weng}, \citenamefont {Qian},\ and\ \citenamefont {Ding}}]{li2019dirac}%
	\BibitemOpen
	\bibfield  {author} {\bibinfo {author} {\bibfnamefont {H.}~\bibnamefont {Li}}, \bibinfo {author} {\bibfnamefont {S.-Y.}\ \bibnamefont {Gao}}, \bibinfo {author} {\bibfnamefont {S.-F.}\ \bibnamefont {Duan}}, \bibinfo {author} {\bibfnamefont {Y.-F.}\ \bibnamefont {Xu}}, \bibinfo {author} {\bibfnamefont {K.-J.}\ \bibnamefont {Zhu}}, \bibinfo {author} {\bibfnamefont {S.-J.}\ \bibnamefont {Tian}}, \bibinfo {author} {\bibfnamefont {J.-C.}\ \bibnamefont {Gao}}, \bibinfo {author} {\bibfnamefont {W.-H.}\ \bibnamefont {Fan}}, \bibinfo {author} {\bibfnamefont {Z.-C.}\ \bibnamefont {Rao}}, \bibinfo {author} {\bibfnamefont {J.-R.}\ \bibnamefont {Huang}}, \bibinfo {author} {\bibfnamefont {J.-J.}\ \bibnamefont {Li}}, \bibinfo {author} {\bibfnamefont {D.-Y.}\ \bibnamefont {Yan}}, \bibinfo {author} {\bibfnamefont {Z.-T.}\ \bibnamefont {Liu}}, \bibinfo {author} {\bibfnamefont {W.-L.}\ \bibnamefont {Liu}}, \bibinfo {author} {\bibfnamefont {Y.-B.}\ \bibnamefont {Huang}}, \bibinfo {author} {\bibfnamefont {Y.-L.}\ \bibnamefont
			{Li}}, \bibinfo {author} {\bibfnamefont {Y.}~\bibnamefont {Liu}}, \bibinfo {author} {\bibfnamefont {G.-B.}\ \bibnamefont {Zhang}}, \bibinfo {author} {\bibfnamefont {P.}~\bibnamefont {Zhang}}, \bibinfo {author} {\bibfnamefont {T.}~\bibnamefont {Kondo}}, \bibinfo {author} {\bibfnamefont {S.}~\bibnamefont {Shin}}, \bibinfo {author} {\bibfnamefont {H.-C.}\ \bibnamefont {Lei}}, \bibinfo {author} {\bibfnamefont {Y.-G.}\ \bibnamefont {Shi}}, \bibinfo {author} {\bibfnamefont {W.-T.}\ \bibnamefont {Zhang}}, \bibinfo {author} {\bibfnamefont {H.-M.}\ \bibnamefont {Weng}}, \bibinfo {author} {\bibfnamefont {T.}~\bibnamefont {Qian}},\ and\ \bibinfo {author} {\bibfnamefont {H.}~\bibnamefont {Ding}},\ }\bibfield  {title} {\bibinfo {title} {Dirac surface states in intrinsic magnetic topological insulators {${\mathrm{EuSn}}_{2}{\mathrm{As}}_{2}$} and {${\mathrm{MnBi}}_{2n}{\mathrm{Te}}_{3n+1}$}},\ }\href {https://doi.org/10.1103/PhysRevX.9.041039} {\bibfield  {journal} {\bibinfo  {journal} {Phys. Rev. X}\ }\textbf {\bibinfo
			{volume} {9}},\ \bibinfo {pages} {041039} (\bibinfo {year} {2019}{\natexlab{c}})}\BibitemShut {NoStop}%
	\bibitem [{\citenamefont {Hao}\ \emph {et~al.}(2019)\citenamefont {Hao}, \citenamefont {Liu}, \citenamefont {Feng}, \citenamefont {Ma}, \citenamefont {Schwier}, \citenamefont {Arita}, \citenamefont {Kumar}, \citenamefont {Hu}, \citenamefont {Lu}, \citenamefont {Zeng}, \citenamefont {Wang}, \citenamefont {Hao}, \citenamefont {Sun}, \citenamefont {Zhang}, \citenamefont {Mei}, \citenamefont {Ni}, \citenamefont {Wu}, \citenamefont {Shimada}, \citenamefont {Chen}, \citenamefont {Liu},\ and\ \citenamefont {Liu}}]{hao2019gapless}%
	\BibitemOpen
	\bibfield  {author} {\bibinfo {author} {\bibfnamefont {Y.-J.}\ \bibnamefont {Hao}}, \bibinfo {author} {\bibfnamefont {P.}~\bibnamefont {Liu}}, \bibinfo {author} {\bibfnamefont {Y.}~\bibnamefont {Feng}}, \bibinfo {author} {\bibfnamefont {X.-M.}\ \bibnamefont {Ma}}, \bibinfo {author} {\bibfnamefont {E.~F.}\ \bibnamefont {Schwier}}, \bibinfo {author} {\bibfnamefont {M.}~\bibnamefont {Arita}}, \bibinfo {author} {\bibfnamefont {S.}~\bibnamefont {Kumar}}, \bibinfo {author} {\bibfnamefont {C.}~\bibnamefont {Hu}}, \bibinfo {author} {\bibfnamefont {R.}~\bibnamefont {Lu}}, \bibinfo {author} {\bibfnamefont {M.}~\bibnamefont {Zeng}}, \bibinfo {author} {\bibfnamefont {Y.}~\bibnamefont {Wang}}, \bibinfo {author} {\bibfnamefont {Z.}~\bibnamefont {Hao}}, \bibinfo {author} {\bibfnamefont {H.-Y.}\ \bibnamefont {Sun}}, \bibinfo {author} {\bibfnamefont {K.}~\bibnamefont {Zhang}}, \bibinfo {author} {\bibfnamefont {J.}~\bibnamefont {Mei}}, \bibinfo {author} {\bibfnamefont {N.}~\bibnamefont {Ni}}, \bibinfo {author} {\bibfnamefont
			{L.}~\bibnamefont {Wu}}, \bibinfo {author} {\bibfnamefont {K.}~\bibnamefont {Shimada}}, \bibinfo {author} {\bibfnamefont {C.}~\bibnamefont {Chen}}, \bibinfo {author} {\bibfnamefont {Q.}~\bibnamefont {Liu}},\ and\ \bibinfo {author} {\bibfnamefont {C.}~\bibnamefont {Liu}},\ }\bibfield  {title} {\bibinfo {title} {Gapless surface {Dirac} cone in antiferromagnetic topological insulator {${\mathrm{MnBi}}_{2}{\mathrm{Te}}_{4}$}},\ }\href {https://doi.org/10.1103/PhysRevX.9.041038} {\bibfield  {journal} {\bibinfo  {journal} {Phys. Rev. X}\ }\textbf {\bibinfo {volume} {9}},\ \bibinfo {pages} {041038} (\bibinfo {year} {2019})}\BibitemShut {NoStop}%
	\bibitem [{\citenamefont {Chen}\ \emph {et~al.}(2019{\natexlab{b}})\citenamefont {Chen}, \citenamefont {Fei}, \citenamefont {Zhang}, \citenamefont {Zhang}, \citenamefont {Liu}, \citenamefont {Zhang}, \citenamefont {Wang}, \citenamefont {Wei}, \citenamefont {Zhang}, \citenamefont {Zuo}, \citenamefont {Guo}, \citenamefont {Liu}, \citenamefont {Wang}, \citenamefont {Wu}, \citenamefont {Zong}, \citenamefont {Xie}, \citenamefont {Chen}, \citenamefont {Sun}, \citenamefont {Wang}, \citenamefont {Zhang}, \citenamefont {Zhang}, \citenamefont {Wang}, \citenamefont {Song}, \citenamefont {Zhang}, \citenamefont {Shen},\ and\ \citenamefont {Wang}}]{chen2019intrinsic}%
	\BibitemOpen
	\bibfield  {author} {\bibinfo {author} {\bibfnamefont {B.}~\bibnamefont {Chen}}, \bibinfo {author} {\bibfnamefont {F.}~\bibnamefont {Fei}}, \bibinfo {author} {\bibfnamefont {D.}~\bibnamefont {Zhang}}, \bibinfo {author} {\bibfnamefont {B.}~\bibnamefont {Zhang}}, \bibinfo {author} {\bibfnamefont {W.}~\bibnamefont {Liu}}, \bibinfo {author} {\bibfnamefont {S.}~\bibnamefont {Zhang}}, \bibinfo {author} {\bibfnamefont {P.}~\bibnamefont {Wang}}, \bibinfo {author} {\bibfnamefont {B.}~\bibnamefont {Wei}}, \bibinfo {author} {\bibfnamefont {Y.}~\bibnamefont {Zhang}}, \bibinfo {author} {\bibfnamefont {Z.}~\bibnamefont {Zuo}}, \bibinfo {author} {\bibfnamefont {J.}~\bibnamefont {Guo}}, \bibinfo {author} {\bibfnamefont {Q.}~\bibnamefont {Liu}}, \bibinfo {author} {\bibfnamefont {Z.}~\bibnamefont {Wang}}, \bibinfo {author} {\bibfnamefont {X.}~\bibnamefont {Wu}}, \bibinfo {author} {\bibfnamefont {J.}~\bibnamefont {Zong}}, \bibinfo {author} {\bibfnamefont {X.}~\bibnamefont {Xie}}, \bibinfo {author} {\bibfnamefont
			{W.}~\bibnamefont {Chen}}, \bibinfo {author} {\bibfnamefont {Z.}~\bibnamefont {Sun}}, \bibinfo {author} {\bibfnamefont {S.}~\bibnamefont {Wang}}, \bibinfo {author} {\bibfnamefont {Y.}~\bibnamefont {Zhang}}, \bibinfo {author} {\bibfnamefont {M.}~\bibnamefont {Zhang}}, \bibinfo {author} {\bibfnamefont {X.}~\bibnamefont {Wang}}, \bibinfo {author} {\bibfnamefont {F.}~\bibnamefont {Song}}, \bibinfo {author} {\bibfnamefont {H.}~\bibnamefont {Zhang}}, \bibinfo {author} {\bibfnamefont {D.}~\bibnamefont {Shen}},\ and\ \bibinfo {author} {\bibfnamefont {B.}~\bibnamefont {Wang}},\ }\bibfield  {title} {\bibinfo {title} {Intrinsic magnetic topological insulator phases in the {Sb} doped {MnBi$_2$Te$_4$} bulks and thin flakes},\ }\href {https://doi.org/10.1038/s41467-019-12485-y} {\bibfield  {journal} {\bibinfo  {journal} {Nat. Commun.}\ }\textbf {\bibinfo {volume} {10}},\ \bibinfo {pages} {4469} (\bibinfo {year} {2019}{\natexlab{b}})}\BibitemShut {NoStop}%
\end{thebibliography}

\end{document}